\input epsf
\documentclass[nohyper,notoc]{article} 
\usepackage{axodraw}
\usepackage{epsfig}

\def\beq{\begin{equation}}
\def\eeq{\end{equation}}
\def\bea{\begin{eqnarray}}
\def\eea{\end{eqnarray}}
\def\bq{\begin{quote}}
\def\eq{\end{quote}}
\def \lsim{\mathrel{\vcenter
    {\hbox{$<$}\nointerlineskip\hbox{$\sim$}}}}
\def \gsim{\mathrel{\vcenter
    {\hbox{$>$}\nointerlineskip\hbox{$\sim$}}}}
\def\gappeq{\mathrel{\rlap {\raise.5ex\hbox{$>$}}
{\lower.5ex\hbox{$\sim$}}}}
\def\lappeq{\mathrel{\rlap{\raise.5ex\hbox{$<$}}
{\lower.5ex\hbox{$\sim$}}}}

\def\a{\alpha}
\def\b{\beta}

\def\s{\sigma}

\def\nnff{(\bar{\nu} \nu)(\bar{f} f)}
\def\llff{(\bar{\ell} \ell)(\bar{f} f)}
\def\lnffp{(\bar{\ell} \nu)(\bar{f} f')}
\def\sw2{\sin^2 \theta_W}
\def\s132{\sin^2 \theta_{13}}

\def \lsim{\mathrel{\vcenter
     {\hbox{$<$}\nointerlineskip\hbox{$\sim$}}}}
\def \gsim{\mathrel{\vcenter
     {\hbox{$>$}\nointerlineskip\hbox{$\sim$}}}}

\newcommand{\as}{^{*)}}

\begin{document}

\renewcommand{\thefootnote}{\fnsymbol{footnote}}

\thispagestyle{empty}
\vspace*{1.cm}

\hfill IFIC/02-36,FTUV-03-0212 

\hfill IPPP/02/49, DCPT/02/98
\vspace{5mm}
\vspace{0.5cm}
\begin{center}
{\large \bf Present and Future Bounds on Non-Standard Neutrino
Interactions}

\vspace{1cm}

{\bf S. Davidson$^{a}$, C. Pe\~na-Garay$^{b}$, N. Rius$^c$,
A. Santamaria$^{c}$}

\vspace{1cm}

{\em (a) IPPP, University of Durham, Durham DH1 3LE,UK }

{\em (b) School of Natural Sciences, Institute for Advanced Study,
Princeton, NJ 08540}

{\em (c) Depto.\ de F\'{\i}sica Te\'orica and IFIC, Universidad de
Valencia-CSIC, Edificio de Institutos de Paterna, Apt. 22085, 46071
Valencia, Spain }

\vspace{1cm}

\begin{abstract}
We consider Non-Standard neutrino Interactions (NSI), described by
four-fermion operators of the form $(\bar{\nu}_{\alpha} \gamma
{\nu}_{\beta}) (\bar{f} \gamma f)$, where $f$ is an electron or first
generation quark.  We assume these operators are generated at
dimension $\geq 8$, so the related vertices involving charged leptons,
obtained by an $SU(2)$ transformation $\nu_{\delta} \rightarrow
e_{\delta}$, do not appear at tree level. These related vertices
necessarily arise at one loop, via $W$ exchange.  We catalogue current
constraints from $\sin^2 \theta_W$ measurements in neutrino
scattering, from atmospheric neutrino observations, from LEP, and from
bounds on the related charged lepton operators. We estimate future
bounds from comparing KamLAND and solar neutrino data, and from
measuring $\sin^2 \theta_W$ at the near detector of a neutrino
factory.  Operators constructed with $\nu_\mu$ and $\nu_e$ should not
confuse the determination of oscillation parameters at a $\nu$factory,
because the processes we consider are more sensitive than oscillations
at the far detector.  For operators involving $\nu_\tau$, we estimate
similar sensitivities at the near and far detector.
\end{abstract}

\end{center}

\section{Introduction}

Lepton flavour violation is observed in atmospheric and solar neutrino
experiments.  The atmospheric neutrino deficit is mainly due to muon
neutrino disappearance: the $\nu_\mu $ flux measured at
Super-Kamiokande (SK) has a strong zenith angle dependence, which
deviates from the Standard Model (SM) expectation by more than
7$\sigma$ \cite{skatm}.  Neutrinos change flavors as they travel to
the Earth from the center of the Sun, as was seen
directly at the Sudbury Neutrino Observatory (SNO) through the
measurements of the charged current and the neutral current reactions
for $^8B$ solar neutrinos \cite{sno}.

These results demonstrate simply that new physics is required. In
principle, the atmospheric and solar neutrino deficits can be
explained by neutrino masses or by giving the neutrinos new
interactions.  Data disfavor non standard interactions (
NSI) of
neutrinos as an explanation for the atmospheric neutrino deficit
\cite{lipari, Fogli:1999fs, Fornengo:2001pm} through the energy and
baseline dependence.  KamLAND detector has recently confirmed the
large mixing angle (LMA) oscillation explanation of the solar neutrino
puzzle\cite{unknown:2002dm}.  Prior to this, NSI were a viable
alternative solution\cite{solarNSI,fitNSI}
\footnote{ Reference \cite{herbi} found that NSI induced in the 
R-parity violating MSSM could not explain the solar neutrino
deficit. However, their operators were otherwise constrained to be at
least an order of magnitude smaller than the solutions found in
\cite{fitNSI}.}.

Neutrino masses are the leading mechanism in the solar and atmospheric
anomalies.  Nonetheless, NSI may be comparable to (or larger
than) oscillation effects in other processes or at other energies.
This is a particularly relevant issue for neutrino factories
\cite{Johnson:1999ci,Ota:2002na,Datta:2000ci,Gago:2001xg,Ota:2001pw,
Huber:2001zw, Huber:2001de,CampanelliRomanino},
where NSI may affect the oscillation parameters inferred from
experimental data, biasing the value of some of the mixing angles. For
instance, it was suggested in \cite{Huber:2001de} that NSI could
interfere with the determination of $\s132$ at a neutrino factory
\footnote{ This interference might be solved through the tail of the
spectra at large energies (Ref.~\cite{CampanelliRomanino}).}. It is
important to understand how significant is this possibility. The
original aim of this paper was to show that short baseline and
precision experiments are more sensitive to NSI than the far detector
at a neutrino factory. We will see that this is true for some NSI, but
borderline for those involving a $\nu_\tau$.

We consider neutral current NSI, from a phenomenological perspective:
we follow \cite{Berezhiani:2001rs}, and assume that the new physics
which induces the non-standard $\nnff$ operator, where $f$ is a
charged lepton ($\ell$) or quark, does $not$ introduce new charged
lepton physics at tree level.
Within this approach, NSI can be constrained
\cite{Berezhiani:2001rs,barger} from neutrino deep inelastic
scattering experiments and from elastic scattering $\nu - e$, where
the baseline is too short for oscillations. NSI would contribute to
$\nu$ scattering events, and therefore to the determination of $\sw2$
in these experiments.  So short baseline, high-flux neutrino
experiments, that measure for instance $\sw2$, can set significant
bounds on NSI involving $\nu_e$ and $\nu_\mu$.

A $\nnff$ operator, will nonetheless induce a $\llff$ and/or $\lnffp$
 operator, where $f'$ is the SU(2) partner of $f$, via external
 one-loop Standard Model dressing.  This is independent of the new
 physics that induces the $\nnff$ operator.  Even in the case in which
 there are no tree level bounds on NSI of neutrinos, because there is
 no appropriate experiment, in general radiative corrections will
 generate other types of interactions which could be tested at present
 (or future) experiments.  Bounds on charged lepton operators
 therefore set model independent bounds on the NSI operators.
 However, at present these bounds are only significant for $\mu - e$
 flavor changing operators, because the loop suppression is a small
 number.  NSI in loops can also make flavour dependent contributions
 to the decay rates of the electroweak gauge bosons, which can set
 relevant bounds on flavour diagonal NSI.

Present bounds still allow sizable (diagonal) NSI of $\nu_\tau$, but
we will show that they can be constrained by comparing SNO/SK solar
data and KamLAND results.  The bounds are significantly better if
KamLAND finds an oscillatory signal, that is if $\Delta m^2_{sol}
\lsim 10^{-4}$ eV, as we will assume for definiteness.

We introduce our notation and assumptions in section 2. Section 3
presents current constraints on flavour diagonal and flavour changing
neutral current NSI, both from tree level effects (short baseline,
high-flux neutrino scattering experiments, LEP, atmospheric neutrinos)
and from one-loop processes (flavour changing charged lepton interactions,
LEP).
In section 4, we discuss the sensitivity of solar neutrino
experiments, SNO and Super-Kamiokande, using KamLAND to pin-point the
solar oscillation parameters, $\Delta m^2_{sol}$ and $\theta_{12}$. We
also discuss the sensitivity of future experiments, using the near
detector at a nufactory as an example.  We summarize our results in
section \ref{summary}, tabulating the best current and future bounds
we obtained.

\section{Notation and Assumptions}

At energy scales $ \ll m_W$ (where there is a large amount of precise
$\nu$ scattering data), the Standard Model interactions of neutrinos
can be described by the effective Lagrangian
\beq
{\cal L}_{eff} = - 2\sqrt{2}G_F ([\bar{\nu}_\beta \gamma_\rho L
\ell_\beta ][\bar{f} \gamma^\rho L f' ] +\hbox{h.c.})  - 2\sqrt{2}G_F
\sum_{P,f,\beta} g^f_{P} [\bar{\nu}_\beta \gamma_\rho L
\nu_\beta][\bar{f} \gamma^\rho P f]
\label{SM}
\eeq
where $P=\{L,R\}$, $\ell$ is a charged lepton, $f$ is a lepton or
quark, $f'$ its SU(2) partner, and the $Z$ couplings $g^f_{P}$ are
given in table~\ref{tab:gAi}.  Greek indices from the beginning of the
alphabet ($\alpha, \beta, $...) label lepton flavours, roman indices
($i,j,$...)  correspond to neutrino mass eigenstates, and late
alphabet Greek letters ($\rho,\sigma$...)  are space-time indices.

We consider non-standard, neutral current neutrino interactions, so we
add operators with the form of the second term in equation
(\ref{SM}). We do not include new charged current interactions. As
discussed in
\cite{Johnson:1999ci,Datta:2000ci,Ota:2001pw,Huber:2001de,g}, NSI can
contribute to a `` neutrino oscillation'' signal via charged current
interactions in the source or detector, or via neutral current
interactions in the propagation from source to detector. However, we
anticipate that other experimental processes are more sensitive to
charged current NSI that long baseline neutrino oscillations ({\it e.g.}
flavour changing NSI in the source could induce taus and wrong sign
muons in the near detector).

\begin{table}
\renewcommand{\arraystretch}{1.45}
$$\begin{array}{|c|cc|}\hline Z\hbox{ couplings} & g^f_L & g^f_R \\
\hline
\nu_e,\nu_\mu,\nu_\tau
& \frac{1}{2} & 0 \\ e,\mu,\tau &-\frac{1}{2}+\sw2 & \sw2 \\ u,c,t &
\frac{1}{2}-\frac{2}{3}\sw2 & -\frac{2}{3}\sw2 \\ d,s,b &
-\frac{1}{2}+\frac{1}{3}\sw2 & \frac{1}{3}\sw2 \\
\hline
\end{array}$$
\caption{$Z$ couplings to SM fermions.
\label{tab:gAi}}
\end{table}
\renewcommand{\arraystretch}{1}

Non-renormalisable operators involving a Standard Model neutrino and
anti-neutrino can be ordered by their dimension, or by their number of
legs in the $SU(3) \times U(1)$ invariant effective theory of SM
fermions and photons.  These options are different, because Higgs
fields saturated by the vacuum expectation value (vev) are not counted
as legs.  If we count by legs, then new physics coupled to neutrinos
can appear as a four fermion operator.  We require this operator
to conserve electric charge, colour and lepton number, which forces it
to be of $V \pm A$ form.  We assume three light neutrinos, with
Majorana masses and the SM interactions of equation (\ref{SM}). We
allow them to have NSI, parametrised as
\beq
{\cal L}_{eff}^{NSI} =- \varepsilon^{fP}_{\alpha \beta} 2 \sqrt{2} G_F
 (\bar{\nu}_\alpha \gamma_{\rho} L \nu_\beta) (\bar{f} \gamma^{\rho}P
 f)
\label{eps}
\eeq
where $f$ is now a first generation SM fermion: $e, u$ or $d$, and $P
= L$ or $R$. We are not concerned with $f$ from the second or third
generation, because such interactions could not affect oscillation
experiments.  We neglect possible CP violation in the new interactions
(this has been considered in
\cite{Gago:2001xg,Gonzalez-Garcia:2001mp}), so we take $
\varepsilon^{fP}_{\alpha \beta} \in \Re$. If the only new neutrino
interactions are neutral current, the neutrino flavour basis is
well-defined (see \cite{g}), and we can label neutrinos by their SM
flavour.

Assuming that the four-fermion vertices of equation (\ref{eps}) arise
in an $SU(2)\otimes U(1)$ gauge invariant theory containing the SM
spectrum with a single Higgs doublet, they can be generated by
operators of dimension six, eight and larger \cite{Berezhiani:2001rs}
containing more and more vevs of the Higgs doublet. In this paper we will
not consider dimension seven $\Delta L$ =2 operators
\cite{Bergmann:2000gn} because, unless their coupling is very small,
we expect these interactions to generate too large neutrino Majorana
masses.  There is an important difference between the dimension 6 and
dimension 8 $(\bar{\nu} \nu)(\bar{f} f)$ operators: new physics which
induces the dimension 6 operator also induces an operator involving
charged leptons, with a coefficient of the same order (by $SU(2)$
invariance)
\cite{Bergmann:1999pk}. Charged lepton physics
imposes tight constraints on these coefficients of dimension 6
operators.  At dimension 8, an operator as in equation (\ref{eps}) can
appear at tree level, for instance as
\beq
\bar {e}_R(H^\dagger \sigma^a \ell)(\bar{\ell} \sigma^a H) e_R
\rightarrow - \frac 1 2
\langle H \rangle^2 (\bar{e} \gamma^{\rho}R e) (\bar{\nu} \gamma_{\rho} 
L \nu)
\label{ex}
\eeq
 without any charged lepton counterpart \cite{Berezhiani:2001rs}
 \footnote{The Lorentz indices in eq. (\ref{ex}) are contracted
 $\bar{\ell} e_R$. The $\sigma$s carry gauge $SU(2)$ indices.}.  We
 include the operator (\ref{eps}), and neglect the associated
 tree-level charged current vertex, so we are effectively assuming
 that these operators are generated at dimension eight.

If new physics operators are generated at tree level by exchange of
particles with mass $\Lambda$, naive power counting tells us that dimension
six operators should give rise to $\varepsilon \sim h^2
v_F^2/\Lambda^2$ where $h$ is some generic coupling of the new physics
particles. Taking into account present data collected at LEP and
Tevatron one can reasonably assume that $\Lambda > 200$~GeV. Even
though somehow lighter particles are not excluded ({\em e.g.} if
neutral or only produced in pairs or having small couplings with
ordinary particles), we are using an effective Lagrangian approach
which is only valid below $\Lambda$. This counting tells us that we
expect $\varepsilon < h^2$ for dimension 6 operators. If NSI
interactions are generated by dimension 8 operators one expects an
extra suppression proportional to $v^2_F/\Lambda^2$ which is only
important if $\Lambda\gg v_F$, therefore one naturally expects
$\varepsilon\ll 1$. However, $h$ could be relatively large without
leaving the perturbative regime. Thus, $\varepsilon$'s order one are
not completely unnatural if the scale of new physics is not extremely
large and the couplings of new particles are large. It is, therefore,
important to check how large these NSI can be on purely
phenomenological grounds.  Thus, in this paper, we will concentrate on
the phenomenology of $U(1) \times SU(3)$ invariant $(\bar{\nu}_\alpha
\nu_\beta) (\bar{f} f)$ operators, with arbitrary coefficients. As
mentioned in the introduction, there will nonetheless be constraints
from charged lepton physics on these operators. One loop SM dressing
of the tree level NSI vertices $(\bar{\nu}_\alpha \nu_\beta) (\bar{f}
f)$ (for instance exchanging a $W$ between the external legs), will
{\it necessarily} induce $(\bar{e}_\alpha e_\beta) (\bar{f} f)$ and/or
$(\bar{\nu}_\alpha e_\beta) (\bar{f} f')$. This is discussed in
section \ref{oneloop}.

In the effective theory described by equations (\ref{SM}) and
(\ref{eps}), there will normally be a number of four-fermion operators
which can contribute to a process.  Experimental measurements of
different processes will set limits on different sums of operators. It
is common, in setting a limit on a given four-fermion operator, to
assume that all other four-fermion operators are zero.  This approach
makes sense when the new physics contributions add incoherently,{\it
c'est \`a dire} when one adds probabilities not amplitudes. It also
makes sense if the different new physics amplitudes interfere with
different SM amplitudes---for instance, if there is an NSI - SM
interference term $\propto
\varepsilon^{qL}_{\alpha \alpha} g_L^q
+ \varepsilon^{qR}_{\alpha \alpha} g_R^q$, these two terms can only
cancel against each other if the NSI know $\sw2$. We consider this
unlikely (since we assume they are generated at dimension 8, rather
than 6), so we quote limits on one $\varepsilon$ at a time in our
summary tables.

We will assume that the only unknown new physics are the non-standard
neutrino interactions. We calculate what current experiments should
measure according to the SM, and require the NSI contribution to be
less than the experimental error (1.6 $\sigma$), or less than the
theory - experiment discrepancy.

\section{Present bounds}
\label{pc}

Before considering particle physics bounds, let us discuss briefly
astrophysical and cosmological bounds on neutrino NSI. Neutrino
interactions with matter, electrons and light quarks, can affect many
astrophysical and cosmological scenarios.  They could keep neutrinos
in thermal equilibrium with ordinary matter for a longer time at the
time of nucleosynthesis and disturb one of the great successes of
modern cosmology. They could also produce stronger interactions of
neutrinos with matter in the core of supernovae therefore keeping
neutrinos trapped for a longer time and disturbing the duration of the
neutrino pulse observed in SN1987A
\cite{raffelt,superNSI,lisiNSI}. They
could also contribute to the energy loss of stars due to processes
like plasmon decay ($\gamma^* \rightarrow \nu \bar{\nu}$) which are
determinant for the evolution of red giants. All these processes occur
in the SM mediated by neutral or charged currents and the SM value is
essential to understand the three scenarios mentioned. Now the
question is how large NSI are allowed in order not to disturb present
observations? For instance, the case of plasmon decay in red giants
has been used to place stringent bounds on a possible neutrino
magnetic moment\cite{raffelt}. The SM gives contributions to plasmon
decay from neutral and charged currents for $\nu_e$ and only from
neutral currents for $\nu_\mu$ and $\nu_\tau$. To destabilize the SM
results one should modify some of these interactions by a factor
larger than one\cite{raffelt}. Therefore, generically one can say that
$\varepsilon< {\cal O}(1)$ from energy loss in red giants. We expect,
at best, similar results from supernova data or nucleosynthesis.  As
we will see, laboratory data already place better limits on $\nu_e$
and $\nu_\mu$ NSI, and SNO, Super-Kamiokande and KamLAND will
also improve the bounds for $\nu_\tau$.

We first consider tree level effects of the operators in (\ref{eps}),
which contain only the neutrino current with either the electron or
first generation quark currents. Low energy scattering experiments can
constrain NSI involving $\nu_e$ and $\nu_\mu$, while to derive bounds
on diagonal $\nu_\tau$ NSI one should use the measurement of the $e^+
e^- \rightarrow \nu \bar\nu \gamma$ cross section at LEP
\cite{Berezhiani:2001rs}
and atmospheric neutrino data \cite{Fornengo:2001pm}.  
We set further bounds using the fact that such operators always induce one loop
effects in much better tested charged lepton processes.  These
constrain flavor changing NSI involving first and second generation
neutrinos to be undetectably small.

\subsection{Tree level effects}

\subsubsection{$\nu$ scattering experiments}

Neutrino NSI with either electrons or first generation quarks can be
constrained by low energy scattering data.  We review previous
analysis \cite{barger, Berezhiani:2001rs} and update them by including
the recent results of the NuTeV experiment.  As we shall see below,
the bounds are rather stringent for $\nu_\mu$ interactions, looser for
$\nu_e$ and do not exist for (diagonal) $\nu_\tau$.

We present bounds assuming that only one operator is present at a
time, for the reasons explained above, although we also comment on how
these limits are relaxed when several diagonal NSI are considered
simultaneously.

\begin{itemize}
\item{ $\nu_e  e \rightarrow \nu e$  scattering}

In the presence of neutral current neutrino NSI the $\nu_e e$ elastic
cross section is given by
\beq
\sigma(\nu_e  e \rightarrow \nu e) = \frac{2 G_F^2 m_e E_\nu}{\pi} \,
\left[(1+g^e_L + \varepsilon^{eL}_{ee})^2
 + \sum_{\alpha \neq e} |\varepsilon^{eL}_{\alpha e}|^2 + \frac 1 3
(g^e_R + \varepsilon^{eR}_{ee})^2 + \frac 1 3 \sum_{\alpha \neq e}
|\varepsilon^{eR}_{\alpha e}|^2
\right]
\label{enue}
\eeq
where $g^e_L=-0.2718$ and $g^e_R= 0.2326$ are the SM neutral current
couplings of the electron, including electroweak radiative corrections
and corresponding to the best fit point of the latest SM global fit of
precision observables (without including NuTeV).

\begin{figure} [t!]
\begin{center}
\epsfig{figure=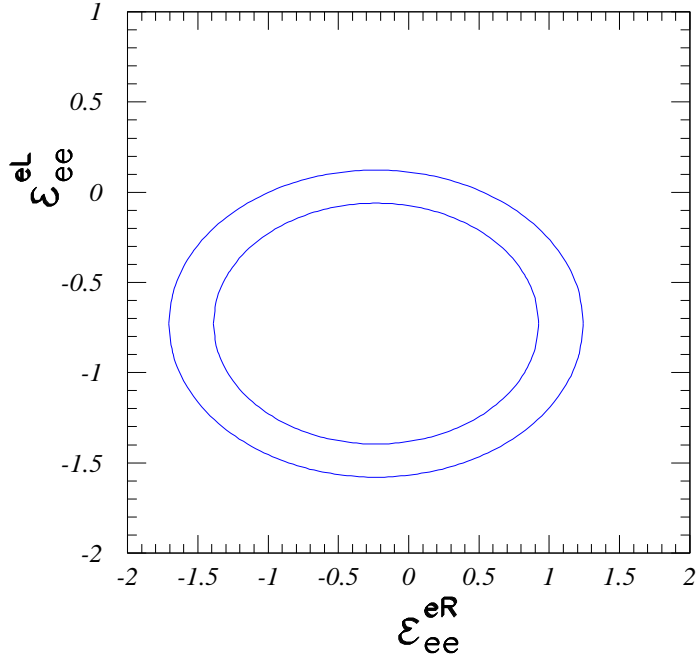,width=.6\textwidth,height=.4\textheight}
\end{center}
\caption{Bounds on flavor conserving non-standard $\nu_e e$ interactions 
from LSND experiment.  Allowed regions at 90\% CL are between the two
ellipses.}
\label{1}
\end{figure}

The most accurate measurement of this cross section is the LSND result
\cite{lsnd}:
\beq
\sigma(\nu_e  e \rightarrow \nu e) = (1.17 \pm 0.17)
\frac{G_F^2 m_e E_\nu}{\pi}  \ ,
\eeq
which, taking into account the SM prediction $\sigma(\nu_e e
\rightarrow \nu_e e)|_{SM}=1.0967 \, G_F^2 m_e E_\nu/\pi$, translates
into the following 90\% CL bounds on diagonal $\nu_e e $ NSI (assuming
only one operator at a time):
\bea
-0.07 < \varepsilon^{eL}_{ee} < 0.11 \\
-1. < \varepsilon^{eR}_{ee} < 0.5
\eea
We can also set bounds on flavour changing NSI. These are only
relevant for $\nu_\tau \nu_e$ interactions, because for $\nu_\mu
\nu_e$ better bounds are obtained from the one loop effects discussed
in section
\ref{oneloop}.
Assuming there are only flavor changing NSI we obtain:
\beq
|\varepsilon^{eL}_{\tau e}| < 0.4
\qquad
|\varepsilon^{eR}_{\tau e}| < 0.7
\eeq

One can wonder how these bounds would be relaxed when allowing for
several operators to be present simultaneously. We consider then both,
left- and right-handed diagonal NSI. The result is shown in
Fig. \ref{1}. The 90\% CL allowed region is between the two ellipses,
and corresponds to the range
\beq
0.445 < (0.7282 + \varepsilon^{eL}_{ee})^2 +
\frac 1 3 \, (0.2326 + \varepsilon^{eR}_{ee})^2 < 0.725
\eeq

\item{ $\nu_e q \rightarrow \nu q$  scattering}

The CHARM collaboration measured the following combination of $\nu_e
N$ and $\bar \nu_e N$ cross sections \cite{charm}:
\beq
R^e=\frac{\sigma(\nu_e N \rightarrow \nu X) +
\sigma(\bar\nu_e N \rightarrow \bar \nu X)}
{\sigma(\nu_e N \rightarrow e X) +
\sigma(\bar\nu_e N \rightarrow \bar{e} X)} =
(\tilde g_{Le})^2 + (\tilde g_{Re})^2 = 0.406 \pm 0.140
\eeq
Since charged current NSI are strongly constrained, we neglect them
and use this measurement to bound the neutrino neutral current NSI.
In this case the effective couplings $(\tilde g_{Le})^2, (\tilde
g_{Re})^2$ are given by
\bea
(\tilde g_{Le})^2 & =& (g^u_L + \varepsilon^{uL}_{ee})^2 +
   \sum_{\alpha \neq e} |\varepsilon^{uL}_{\alpha e}|^2 + (g^d_L +
   \varepsilon^{dL}_{ee})^2 + \sum_{\alpha \neq e}
   |\varepsilon^{dL}_{\alpha e}|^2 \label{gl2} \\ (\tilde g_{Re})^2&=
   &(g^u_R + \varepsilon^{uR}_{ee})^2 + \sum_{\alpha \neq e}
   |\varepsilon^{uR}_{\alpha e}|^2 + (g^d_L + \varepsilon^{dR}_{ee})^2
   + \sum_{\alpha \neq e} |\varepsilon^{dR}_{\alpha e}|^2 \ .
\label{gr2}
\eea

The SM couplings corresponding to the best fit are $(\tilde
g_{Le})^2=0.3042 $ and $(\tilde g_{Re})^2=0.0301$.  Using this result,
the 90\% CL bounds on flavour diagonal NSI are
\bea
-1. < \varepsilon^{uL}_{ee} < 0.3
\\
-0.3< \varepsilon^{dL}_{ee} < 0.3
\\
-0.4 < \varepsilon^{uR}_{ee} < 0.7
\\
-0.6 < \varepsilon^{dR}_{ee} < 0.5
\eea
when assuming only one operator at a time.  

The corresponding 90\% CL bounds for flavour changing NSI interactions
are
\beq
|\varepsilon^{qP}_{\tau e}| < 0.5
\qquad q=u,d \qquad P=L,R
\eeq
Again, these bounds are only relevant for $\nu_e \nu_\tau$, since for
$\nu_e \nu_\mu$ tighter ones are derived from one loop effects.

If we consider all kind of diagonal NSI, the allowed regions at 90\%
CL are limited by two four dimensional ellipsoids and are given by
\beq
0.176 < (0.3493 + \varepsilon^{uL}_{ee})^2 + (-0.4269 +
\varepsilon^{dL}_{ee})^2 + (-0.1551 + \varepsilon^{uR}_{ee})^2 +
(0.0775 + \varepsilon^{dR}_{ee})^2 < 0.636
\eeq

\item{$\nu_\mu e \rightarrow \nu e$}

The CHARM II collaboration gives the following results for vector and
axial-vector $e-\nu_\mu$ couplings \cite{charmII}:
\beq
g^e_V= -0.035 \pm 0.017
\qquad\hbox{and}\qquad
g^e_A= -0.503 \pm 0.017
\eeq
where they have used LEP forward-backward asymmetry to determine the
signs.  From these one gets
\beq
g^e_L= -0.269 \pm 0.017
\qquad\hbox{and}\qquad
g^e_R= 0.234 \pm 0.017
\eeq
The SM values of the left- and right-handed couplings are the same as
for $\nu_e e $ scattering, so we can readily derive the 90\% CL bounds
on diagonal NSI
\bea
-0.025 < \varepsilon^{eL}_{\mu \mu} < 0.03
\\
-0.027 < \varepsilon^{eR}_{\mu \mu} < 0.03
\eea
as well as on flavour changing operators,
\beq
|\varepsilon^{eP}_{\tau \mu}| < 0.1
\qquad P=L,R
\eeq
when we allow only flavor changing NSI.

\item{$\nu_\mu q \rightarrow \nu q$}

The NuTeV collaboration measures the ratios of neutral current to
charged current neutrino-nucleon cross sections, which for an
isoscalar target and at leading order are given by
\begin{eqnarray}
R_\nu &\equiv& \frac{\sigma(\nu N \to \nu X)}{\sigma(\nu N\to \mu X)}
=
(\tilde g_{L\mu})^2 + r (\tilde g_{R\mu})^2\\ 
R_{\bar{\nu}} &\equiv&
\frac{\sigma(\bar\nu N \to \bar\nu X)}{\sigma(\bar\nu  N\to \bar\mu X)} =
(\tilde g_{L\mu})^2 + \frac{1}{r} (\tilde g_{R\mu})^2 \ ,
\end{eqnarray}
where
\beq
r=\frac{\sigma(\bar\nu N\to \bar\mu X)}{\sigma(\nu N\to \mu X)} \ .
\label{eqnr}
\eeq
Neglecting charged current NSI, the effective couplings $(\tilde
g_{L\mu})^2, (\tilde g_{R\mu})^2$ are as given in
eqs.(\ref{gl2}),(\ref{gr2}), just changing $e \to \mu$ in the
coefficients of the neutral current NSI.
The values of these couplings reported by NuTeV are
\cite{nutev}
\beq
(\tilde g_{L\mu})^2 = 0.3005 \pm 0.0014\qquad\hbox{and}\qquad (\tilde
g_{R\mu})^2 = 0.0310 \pm 0.0011 \ .
\eeq
While $(\tilde g_{R\mu})^2$ is in agreement with the SM, $(\tilde
g_{R\mu})^2_{SM}=0.0301$, $(\tilde g_{L\mu})^2$ is about 3$\sigma$
away from the SM prediction, $(\tilde g_{L\mu})^2_{SM}=0.3042$.

The NuTeV result for $(\tilde g_{L\mu})^2$ can be fitted (at 90\% CL)
by
\beq
-0.009 < \varepsilon^{uL}_{\mu \mu} < -0.003
\qquad\hbox{or}\qquad
0.002 < \varepsilon^{dL}_{\mu \mu} < 0.008 \ ,
\eeq
and, in principle, since the measured $(\tilde g_{L\mu})^2$ is smaller
than the SM prediction, pure left-handed flavour changing NSI are
excluded because they do not interfere with the SM amplitude and
therefore always give a positive contribution.

Alternatively one can assume some other explanation
\cite{Davidson:2001ji} of the discrepancy, and estimate
that NSI should contribute less than 1.64 $\sigma$ to the NuTeV
result.  This leads to the constraints:
\bea 
|\varepsilon^{qL}_{\mu \mu}| < 0.003 & q=u,d \\ |\varepsilon^{qL}_{\tau
\mu}| < 0.05 & q=u,d
\eea

\begin{figure} [t!]
\begin{center}
\epsfig{figure=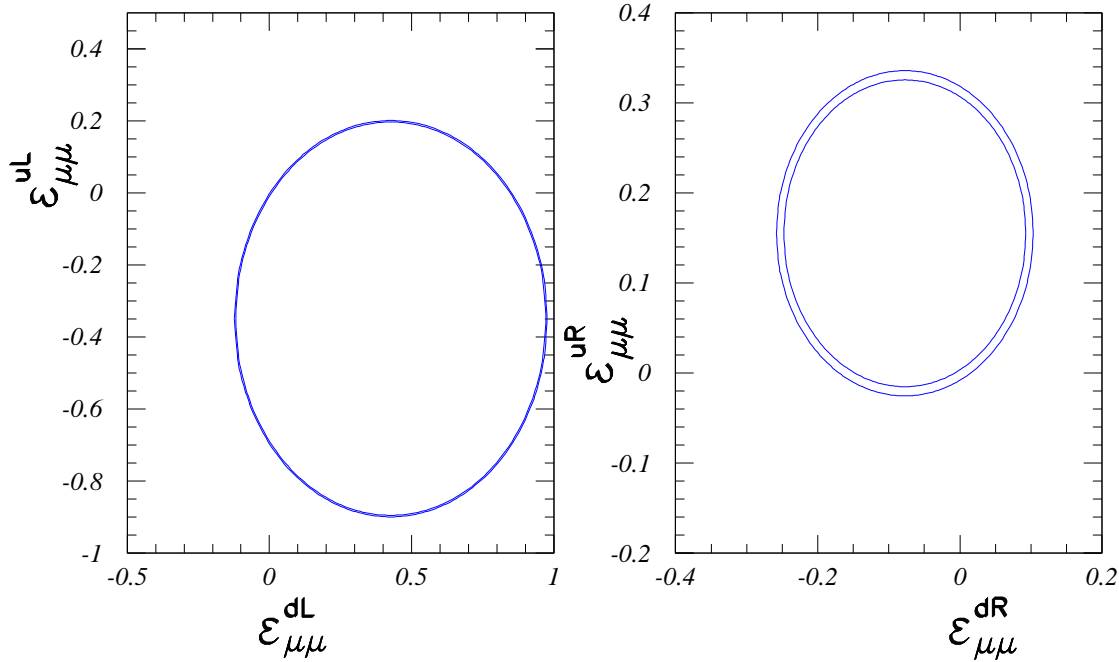,width=.9\textwidth,height=.4\textheight}
\end{center}
\caption{Bounds on flavor conserving non-standard $\nu_\mu q$ interactions 
from NuTeV experiment. The 90\% CL allowed regions are the thick
ellipse in the left panel and the region between the two ellipses in
the right one.}
\label{2}
\end{figure}

For diagonal right-handed NSI the 90\% CL allowed ranges are
\bea
-0.008 < \varepsilon^{uR}_{\mu \mu} < 0.003
\\
-0.008 < \varepsilon^{dR}_{\mu \mu} < 0.015 \ ,
\eea
while for flavour changing interactions,
\beq
|\varepsilon^{qR}_{\tau \mu}| < 0.05 \ .
\qquad q=u,d
\eeq

These bounds are relaxed if we consider several operators present
simultaneously. In Fig. \ref{2} we plot the 90\% CL limits on diagonal
neutrino NSI with u- and d-type quarks, both for left- and
right-handed interactions. The allowed regions are the thick ellipse
in the left panel
\footnote{There are indeed two ellipses, but they can not be distinguished
in the figure.} and the region between the two ellipses in the right
one. They are described by the equations:
\bea
0.2982 < & (0.3493 + \varepsilon^{uL}_{\mu \mu})^2 + (-0.4269 +
\varepsilon^{dL}_{\mu \mu})^2 & < 0.3028
\\
0.0292 < & (-0.1551 + \varepsilon^{uR}_{\mu \mu})^2 + (0.0775 +
\varepsilon^{dR}_{\mu \mu})^2 & < 0.0328
\eea

\end{itemize}

\subsubsection{LEP }

The authors of Ref. \cite{Berezhiani:2001rs} have pointed out the
importance of the $e^+ e^- \rightarrow \nu \bar{\nu} \gamma$ cross
section measured at LEP II in order to constraint neutrino NSI.  For
the case of diagonal $\nu_\tau e$ interactions these are the only
laboratory bounds, and for $\nu_e e$ they are comparable to the LSND
limits already discussed. We refer the reader to
\cite{Berezhiani:2001rs} for the detailed analysis, and just summarize
here the 90\% CL bounds when only $\nu_\tau e$ NSI are considered,
which can be read from their Fig. 4:
\footnote{We have estimated a $\sim 10 \%$ accuracy at 90\% CL.}
\bea
-0.6 < \varepsilon^{eL}_{\tau \tau} < 0.4
\\
-0.4 < \varepsilon^{eR}_{\tau \tau} < 0.6
\eea

This reaction is also useful to constrain flavor changing NSI, but the
bounds are comparable or looser than the ones derived from elastic
scattering, namely
\beq
|\varepsilon^{eP}_{\alpha \beta}| < 0.4
\qquad P=L,R,  \ \alpha=\tau, \ \beta=e, \mu
\eeq
assuming only flavor changing non-standard operators.

 This process cannot constrain $\nu_\tau q $ interactions. However, it
is obvious that any NSI interaction can contribute to $Z\rightarrow f
\bar{f}
\nu \bar{\nu}$ at LEP1 or $e e^+ \rightarrow f\bar{f} \nu \bar{\nu}$ at LEP2.
These processes occur in the SM with a pair of virtual $Z, W, \gamma$.
They have also been observed and the observations roughly agree with
the SM expectations. For instance the cross section with virtual $Z$
and $\gamma^*$ going to quarks and neutrinos $e^+ e^- \rightarrow Z
\gamma^* \rightarrow q q \nu \nu$ has been measured by DELPHI at
energies above 189 GeV to be $\sigma_{Z\gamma^*}=(0.129\pm 0.038)$
which has to be compared with the SM prediction of about
$0.092$--$0.084$.  This suggest that NSI should be at most as strong
as SM interactions. It is difficult to extract more precise
information on possible NSI from these data because it is based on a
particular pole structure which is not shared by the NSI. Clearly, to
be more precise on the NSI a dedicated study should be performed.  In
addition, NSI of $\nu_\tau$ with quarks can in principle be
constrained through one loop effects both in the invisible and the
hadronic Z width, but typically these bounds are ${\cal O}(1)$ (see
sect.~\ref{oneloop}).

 One can ask whether TESLA would set improved bounds.  The sensitivity
at TESLA to contact interactions of the form $\eta 2 \sqrt{2} G_F
(\bar{e} \gamma^\rho P e)(\bar{\tau}
\gamma_\rho \mu,e)$ can be estimated \cite{Cuypers:1996ia}
to be $\eta \gsim 0.5 \times 10^{-3}$. Using $c \sim .002$ from
section \ref{oneloop}, $ \eta \sim c \varepsilon$ implies $\varepsilon
< O(0.3)$.  This is not particularly significant, and corresponds to a
new physics mass scale below 500 GeV ($\sqrt{s}$ at TESLA) so the
contact interaction analysis is not appropriate. A more promising
channel might be $e \bar{e} \rightarrow \gamma (\nu \bar{\nu})$, but
to our knowledge, the limits that could be extracted from searching
for this at TESLA have not been studied.

\subsubsection{Atmospheric neutrinos}

Exotic (i.e., no-oscillation) solutions of the atmospheric neutrino
problem are disfavored by the energy and baseline dependence
\cite{lipari,Fogli:1999fs}.
In particular, in Ref.\cite{Fornengo:2001pm} an explanation of
atmospheric neutrino data in terms of pure NSI of neutrinos is ruled
out at 99\% CL, but a combined analysis is performed which includes
both, oscillations and non-standard neutrino-matter interactions, and
allows to set significant bounds on diagonal as well as flavour
changing neutrino's NSI.  Two comments are in order. First, note that
such bounds apply to the vector coupling constant of the NSI,
$\varepsilon^{fV}_{\alpha \beta}=
\varepsilon^{fL}_{\alpha \beta} + \varepsilon^{fR}_{\alpha \beta}$,
since it is the only one which appears in neutrino propagation through
matter. Second, the study does not consider NSI in the production
neither in the detection of the neutrinos. However both processes take
place through charged current interactions, which are better
constrained, so one does not expect sizable effects.

Notice that when NSI appear only in the propagation of neutrinos in
matter, the relevant parameters are the flavour changing NSI couplings
and the {\em difference} between the diagonal ones, denoted in
\cite{Fornengo:2001pm,Maltoni:2002kq} as the non-universality parameter 
$\varepsilon'$.
The combined analysis leads to the following 90\% CL bounds when both
diagonal and flavour changing NSI are simultaneously present
\cite{Maltoni:2002kq}:
\beq
-0.016 < \varepsilon^{dV}_{\mu \tau} < -0.009 
\qquad 0 < \varepsilon^{dV}_{\mu \tau}< 0.013
\eeq
\beq
|\varepsilon^{dV}_{\mu \mu} - \varepsilon^{dV}_{\tau \tau}| < 0.03
\eeq
Although the fit has been done assuming only neutrino-down quark NSI
similar results can be expected for the neutrino-electron and
neutrino-up quark cases.

This result leads to a stringent constraint on flavour changing
four-fermion operators involving $\nu_\mu \nu_\tau$, and provides a
complementary bound on diagonal NSI of $\nu_\tau$ with quarks, only
loosely constrained by LEP.

\subsection{One loop effects}
\label{oneloop}

On general grounds we expect that interactions in which the \( \nu
_{\alpha} \) are replaced by the corresponding leptons will be
generated by one loop diagrams with virtual \( W \)'s. Effective
interactions, however, are nonrenormalizable and, therefore, a precise
prescription has to be given in order to estimate these
corrections. Our point of view will be that these are originated from
a more complete theory at scales \( \Lambda \gg m_{W} \) which is
renormalizable (or perhaps finite) and in which observables can be
computed in terms of a few parameters.

We illustrate our point with a simple toy model, in which exact loop
calculations can be easily done. It is {\it not} a realistic example
of the type of model we wish to constrain (see {\it e.g.}
\cite{Berezhiani:2001rs}  for such models), because it induces
$\left( \bar{e}\gamma ^{\rho }Pe\right)
\left( \bar{\nu }_{\mu}\gamma _{\rho }L\nu _{\tau} \right) $
and $\left( \bar{\tau}\gamma ^{\rho }P e\right)
\left( \bar{\nu }_{e}\gamma _{\rho }L\nu _{\mu} \right) $
simultaneously at dimension 6 with approximately the same
coefficient. So $\tau$ appearance at the near detector of a neutrino
factory would be more sensitive to this model than neutrino
scattering.

As a guide for the type of calculations we are going to perform, we
imagine the standard model extended by a singly charged scalar singlet
\( h^{+} \), with mass $M$, which has the following interaction with
the standard leptonic doublet, \( \ell \), (\( \tilde{\ell }=i\tau
_{2}\ell ^{c} \))
\begin{equation}
\label{eq:singlet-lagrangian}
\mathcal{L}_{h}=f_{\alpha\beta}\left( \overline{\ell _{\alpha}}
\tilde{\ell }_{\beta}\right) h^{-}+\mathrm{h.c.}~,
\end{equation}
where \( f_{\alpha\beta} \) is a coupling antisymmetric in flavour
indices. Expanding in the fields \begin{equation}
\label{eq:singlet-lagrangian-bis}
\mathcal{L}_{h}=2f_{\alpha\beta}\left( \overline{\nu _{L\alpha}}e_{L\beta}^{c}
\right) h^{-}+\mathrm{h.c.}~.
\end{equation}
Exchange of scalars will generate a four-fermion interaction of the
type we are considering (for a review of this model from the effective
Lagrangian point of view see~\cite{Bilenky:1993bt}),
\[
\mathcal{L}=4f_{\alpha \beta}f_{\gamma \delta}^{*}\frac{1}{M^{2}}
\left( \overline{e^{c}_{L\delta}}\nu _{L\gamma}\right) 
\left( \bar{\nu }_{L\alpha}e^{c}_{L\beta}\right) =
2f_{\alpha \beta}f_{\gamma \delta}^{*}\frac{1}{M^{2}}
\left( \overline{e_{\beta}}\gamma ^{\rho }Le_{\delta}\right) 
\left( \bar{\nu }_{\alpha}\gamma _{\rho}L\nu _{\gamma}\right)~. \]
We are only interested in NSI with electrons, so we take
\( f_{\mu \tau }=0 \) and find that this model gives
\begin{equation}
\label{eq:NSI-FV-singlet}
\mathcal{L}=-2\sqrt{2}G_{F}\varepsilon ^{eL}_{\mu \tau }
\left( \overline{e}\gamma ^{\rho }Le\right) \left( \bar{\nu }_{\mu }
\gamma _{\rho }L\nu _{\tau }\right) +\mathrm{h.c.}+\cdots~,
\end{equation}
where
\begin{equation}
\label{eq:epsilon-singlet}
\varepsilon ^{eL}_{\mu \tau }=-\frac{f_{\mu e}f_{\tau e}^{*}}{g^{2}}4
\frac{m^{2}_{W}}{M^{2}} ~.
\end{equation}
In addition this model also generates, at tree level, interactions of
the type
\( \left( \overline{e}\gamma ^{\rho}L\tau \right) 
\left( \bar{\nu }_{\mu }\gamma _{\rho }L\nu _{e}\right)  \)
and \( \left( \overline{\mu }\gamma ^{\rho}L\tau \right)
\left( \bar{\nu }_{e}\gamma _{\rho }L\nu _{e}\right)  \)
which are not interesting for our discussion and which we have
represented by the dots in eq.~(\ref{eq:NSI-FV-singlet}).

\begin{figure}
{\par\centering \includegraphics{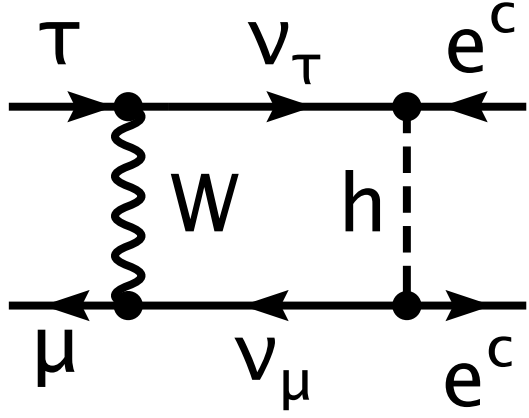} \par}
\caption{One-loop contributions to four-fermion interactions in a theory with a charged
scalar singlet.\label{fig:box}}
\end{figure}

\begin{figure}
{\par\centering
\resizebox*{.55\textwidth}{!}{\includegraphics{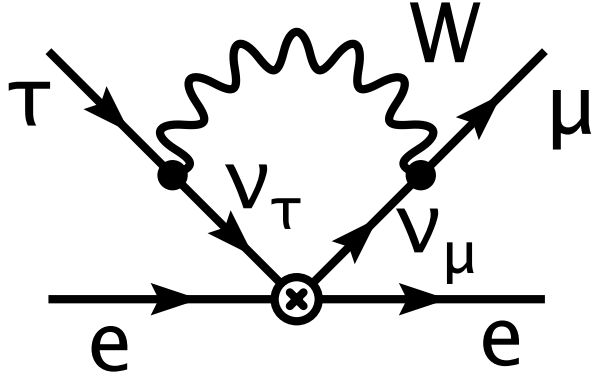}} \par}
\caption{One-loop contributions to four-fermion interactions in the effective 
theory.\label{fig:boxEFT}}
\end{figure}

Although at tree level this model does not provide any contribution to
\( \tau ^{-}\rightarrow \mu ^{-}e^{-}e^{+} \) it does contain
contributions at the one-loop level. The only contribution with
\( \nu _{\mu } \) and \( \nu _{\tau } \) in virtual states is given by the
diagram in Fig.~\ref{fig:box} (note that in this model there are other
diagrams contributing to the process with other types of neutrinos in
virtual states).
 In fact the diagram in Fig.~\ref{fig:box} can be easily computed and it is
finite. In the limit of \( M_{W},M\gg m_{\tau } \) it generates the
following interaction among four charged leptons (after Fierz
reordering)
\begin{equation}
\label{eq:tau-mu-ee-singlet}
\mathcal{L}=-2\sqrt{2}G_{F}\varepsilon ^{eL}_{\mu \tau }\frac{\alpha }{8\pi
s^{2}_{W}}F\left( \frac{M^{2}}{m^{2}_{W}}\right)
\left( \overline{e}\gamma^{\rho }Le\right) 
\left( \overline{\mu }\gamma _{\rho }L\tau \right) \, ,
\end{equation}
where \( \varepsilon ^{eL}_{\mu \tau } \) is given by
eq.~(\ref{eq:epsilon-singlet}) and the function \( F(x) \) is
\[
F(x)=\frac{x}{x-1}\ln (x)\approx \ln (x)\, .
\]
The important thing is that to obtain this logarithmic contribution we
do not need to know all the details of the complete theory. We can
compute it by using the four-fermion effective interaction
(Fig.~\ref{fig:boxEFT}).

This diagram yields a logarithmically divergent contribution.
However, the divergent part can be unambiguously calculated. In the
effective field theory language one should add a counter-term (a
4-charged fermion interaction) to the Lagrangian in order to absorb
this divergence. The coefficient of this new operator will run under
the influence of the diagram in Fig.~\ref{fig:boxEFT}. Therefore even
if at the scale of new physics, \( \Lambda \), there is no 4-charged
fermion interaction at tree level it will be generated through running
from the scale \( \Lambda \) to the electroweak scale. Thus, to
compute the logarithmic contribution it is enough to compute the
divergent part of the diagram in Fig.~\ref{fig:boxEFT}.  The finite
parts have to do with possible non-logarithmic contributions generated
at the scale \( \Lambda \) which cannot be computed without knowing
the details of the complete theory and should be absorbed in the
initial conditions for the running of the different
operators. However, if the scale \( \Lambda \) is much larger than the
electroweak scale, one can reasonably assume that the logarithm \( \ln
(\Lambda/m_{W}) \) dominates completely the result.

If we apply this point of view to the model we just considered we
obtain exactly the same result as in eq.~(\ref{eq:tau-mu-ee-singlet})
with \( F(M^{2}/m^{2}_{W}) \) replaced by \( 2\ln
(\Lambda/m_{W})+\kappa (\Lambda ) \), where \( \kappa (\Lambda ) \)
takes into account possible non-logarithmic contributions at the scale
\( \Lambda \).  So the effective Lagrangian calculation gives the
correct answer if we identify
\( \Lambda  \) with the mass of the charged singlet, \( M \) and take \( \kappa (M)=0 \).
Note, however, that in the effective field theory calculation we have
no way to determine \( \kappa (\Lambda ) \), which depends on the
details of the complete theory. In fact, in the effective field theory
\( \Lambda \) can only loosely be related to the masses of the unknown
more complete theory. \( \kappa (\Lambda ) \), somehow, parameterizes
all these unknown details. The important point is that we expect it to
be at most order 1 and negligible in front of the calculable
logarithmic piece if \( \Lambda \) is large enough.

In order to set reliable bounds on the \( \varepsilon \)'s using these
one loop corrections we would need to know roughly the size of \(
\Lambda \).  However this is not known. Bounds will be set directly on
the \( \varepsilon \)'s at tree level and on \( \varepsilon \ln
(\Lambda/m_{W}) \) at one loop. Of course, \( \varepsilon \) also
depends implicitly on \( m^{2}_{W}/\Lambda ^{2} \) as in
eq.~(\ref{eq:epsilon-singlet}), however, it also depends on other
parameters, in this case the Yukawa and gauge couplings, which, at the
level of the effective four-fermion theory cannot be completely
disentangled. One might expect the logarithm to give some enhancement,
but, since the size of this enhancement cannot be reliably computed we
choose to be conservative and take \( \ln (\Lambda/m_{W})\approx 1 \)
in all the bounds we will set from loop calculations.

\subsubsection{Limits on lepton flavor violating interactions}

Using these arguments we can get some indirect bounds on the
interactions in eq.~(\ref{eps}) by using radiative corrections. The
calculation of the diagram in Fig.~\ref{fig:boxEFT} remains
essentially unchanged if we use as starting point an effective
interaction with other type of neutrinos or with neutrinos and quarks,
thus, if we have an interaction among electrons, u-quarks or d-quarks
with neutrinos with strengths
\( \varepsilon _{\alpha\beta}^{eP} \), \( \varepsilon _{\alpha\beta}^{uP} \), \( \varepsilon _{\alpha\beta}^{dP} \)
respectively, we expect a four-fermion interaction among electrons,
u-quarks or d-quarks with the charged leptons with strengths \(
c\varepsilon _{\alpha\beta}^{eP} \), c\( \varepsilon
_{\alpha\beta}^{uP} \), c\( \varepsilon _{\alpha\beta}^{dP} \) with
\begin{equation}
\label{eq:c}
c=\frac{\alpha }{4\pi s^{2}_{W}}\ln \left(
\frac{\Lambda}{m_{W}}\right)
\approx 0.0027\, .
\end{equation}
These interactions give rise to a class of interesting processes like
\( \mu ^{-}\rightarrow e^{+}e^{-}e^{-} \) ($BR < 1.0 \times
10^{-12}$),
\( \tau ^{-}\rightarrow e^{+}e^{-}e^{-} \) 
($BR < 2.9 \times 10^{-6}$), \( \tau ^{-}\rightarrow e^{+}e^{-}\mu
 ^{-} \) ($BR < 1.7 \times 10^{-6}$),
\( \mu \, Ti\rightarrow e\, Ti \)($\Gamma < 4.3 \times
10^{-12}$), \( \tau ^{-}\rightarrow e^{-}\pi ^{0} \)($BR < 3.7 \times
 10^{-6}$),
\( \tau ^{-}\rightarrow \mu ^{-}\pi ^{0} \)($BR < 4.0 \times
10^{-6}$), \( \tau ^{-}\rightarrow e^{-}\rho ^{0} \)($BR < 2.0 \times
 10^{-6}$),
\( \tau ^{-}\rightarrow \mu ^{-}\rho ^{0} \)($BR < 6.3 \times
10^{-6}$) which are strongly bounded from present experiments, thus
one can obtain some information on \( \varepsilon _{e\mu }^{eP} \),
\( \varepsilon _{e\tau }^{eP} \), \( \varepsilon _{\mu \tau }^{eP} \), and
similarly for quark interactions with neutrinos.  We have listed the
branching ratios we use to set bounds, so that the limits on the
$\varepsilon$'s can be rescaled if the experimental bounds become
stronger.

Consider for example the bound on $\varepsilon_{\tau \mu}^{eP}$ from
$\tau^-\rightarrow \mu^- e^+ e^-$. The diagram of
Fig.~\ref{fig:boxEFT} with the vertex $\otimes$ given by
eq.~(\ref{eps}) provides an interaction (like that in
eq.~(\ref{eq:tau-mu-ee-singlet}))
\begin{equation}
\label{eq:tau-mu-ee}
\mathcal{L}=-2\sqrt{2}G_{F} c \varepsilon ^{eP}_{\mu \tau }
\left( \overline{e}\gamma ^{\rho }Pe\right) 
\left( \overline{\mu }\gamma _{\rho}L\tau \right) \, ,
\end{equation}
with $c$ given in eq.~(\ref{eq:c}). From this we can immediately
compute the branching ratio
\begin{equation}
\label{eq:br-tau-mu}
BR(\tau^-\rightarrow\mu^- e^+ e^-) = BR(\tau^-\rightarrow\mu^-
\nu_\tau\bar{\nu}_\mu)
\frac{\Gamma(\tau^-\rightarrow\mu^- e^+ e^-)}
{\Gamma(\tau^-\rightarrow\mu^-\nu_\tau\bar{\nu}_\mu)}= 0.1737 \left|c
\varepsilon_{\tau \mu}^{eP}\right|^2<1.7\times 10^{-6}\ ,
\end{equation}
from where we obtain
\begin{equation}
\label{eq:bound-epsilon-tau-mu}
\left|\varepsilon_{\tau \mu}^{eP}\right|< 0.0031/c <1.2\, , 
\quad 90\%\ \ \mathrm{CL}~. 
\end{equation}

Note that, as can easily be checked from the Michel parameters, the
operators $(\bar{\ell}_\beta \gamma^\rho \ell_\alpha ) ( \bar{e}
\gamma_\rho L e)$ and $(\bar{\ell}_\beta \gamma^\rho \ell_\alpha )$ $(
\bar{e} \gamma_\rho R e)$ give the same contribution to the total
decay rate.

Somehow worse bounds can be obtained for $\varepsilon_{\tau e}^{eP}$,
$|\varepsilon_{\tau e}^{eP}|<2.9$, because the limits on
$\tau^-\rightarrow e^-e^+ e^-$ are a bit looser and because of the two
identical particles in the final state.

Much more interesting, however, are the bounds that can be set on
$\varepsilon_{\mu e}^{eP}$ because of the strong experimental limit on
$BR(\mu^-\rightarrow e^- e^+ e^-)$. In this case the calculation is
similar except for a few factors due to the identical particles in the
final state.  We have
\begin{equation}
\label{eq:br-tau-e}
BR(\mu^-\rightarrow e^- e^+ e^-) \approx
\frac{\Gamma(\mu^-\rightarrow e^- e^+ e^-)}
{\Gamma(\mu^-\rightarrow e^-\nu_\mu\bar{\nu}_e)}= 0.5\left|c
\varepsilon_{\mu e}^{eP}\right|^2<1\times 10^{-12}\ ,
\end{equation}
from where we obtain
\begin{equation}
\label{eq:bound-epsilon-tau-e}
\left|\varepsilon_{\mu e}^{eP}\right|< 1.4\times 10^{-6}/c <5\times10^{-4}\, ,
\quad 90\%\ \ \mathrm{CL}~.
\end{equation}

Similar arguments can be used for NSI of quarks with neutrinos. In
fact, Fig.~\ref{fig:boxEFT} also generates interactions like
eq.~(\ref{eq:tau-mu-ee-singlet}) but with the electron fields replaced
by $u$ or $d$-quark fields. These interactions contribute to several
hadronic decays of the tau lepton. Thus, we can set bounds on the
$\varepsilon^{qP}_{\tau\beta}$ (with $\beta=e,\mu$ and $q=u,d$).  For
instance, assuming no unnatural cancellations among the $u$-quark and
the $d$-quark or the $P=L$ and $P=R$ contributions, we have
\beq
|\varepsilon^{qP}_{\tau e}| < \sqrt{2}/c\sqrt{\frac{ BR( \tau^-
\rightarrow e^- \rho^0)}{ BR( \tau^- \rightarrow
\nu_\tau \rho^-)}}<1.6~,
\eeq
where we included the isospin factor $\sqrt{2}$ and used $BR( \tau^-
\rightarrow \nu_\tau \rho^-) \approx 0.22$.

From the decay $\tau^-\rightarrow e^- \pi^0$ one obtains similar but
slightly worse bounds because the limit on $BR( \tau^- \rightarrow e^-
\pi^0)$ is worse and because $BR( \tau^- \rightarrow
\nu_\tau \pi^-)$ is smaller. It is important, however, to remark that decays
into $\rho$'s probe the vector channel and decays into $\pi$'s probe
the pseudoscalar channel and, in this sense, provide a complementary
information.

Using $\tau^- \rightarrow \mu^- \rho^0$ (or $\tau^- \rightarrow \mu^-
\pi^0$) one can also set bounds on
$\varepsilon^{qP}_{\tau\mu}$. Using decays into $\rho$'s we 
obtain $|\varepsilon^{qP}_{\tau\mu}| < 2.8$.

Finally we can set bounds on $\varepsilon_{\mu e}^{qP}$ from $\mu-e$
conversion on nuclei.
Comparing the experimental upper bound on the rate of $\mu-e$
conversion to the rate of muon capture,
\[
R_{\mu e}\equiv \frac{\sigma(\mu^- Ti \rightarrow e^- Ti)}
{\sigma(\mu^- Ti \rightarrow \mathrm{capture})}< 4.3\times 10^{-12} ~,
\]
implies (see {\it e.g.} \cite{deGouvea:2000cf}) that
\[
\varepsilon^{qP}_{\mu e} \lsim \sqrt{R_{\mu e}}/c \approx 7.7\times 10^{-4}~.
\]
We have collected the numerical values of the best bounds in
table~\ref{tab:fcffv}.

\subsubsection{Limits on lepton flavor conserving interactions}

\begin{figure}
{\par\centering\resizebox*{.5\textwidth}{!}{\includegraphics{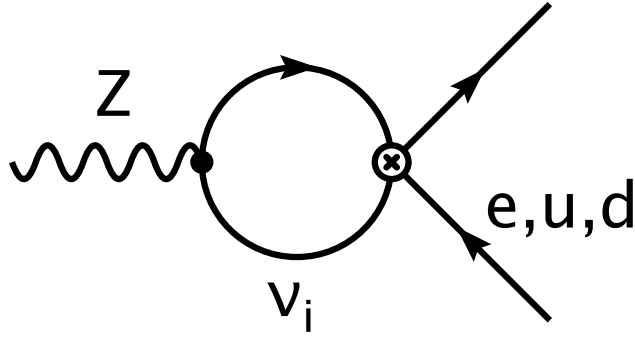}}
\par}
\caption{Contributions of lepton flavor conserving NSI to the vertex of the 
Z-gauge bosons to electrons, $u$ and $d$-quarks.\label{fig:Z-eud}}
\end{figure}

\begin{figure}
{\par\centering\resizebox*{.5\textwidth}{!}{\includegraphics{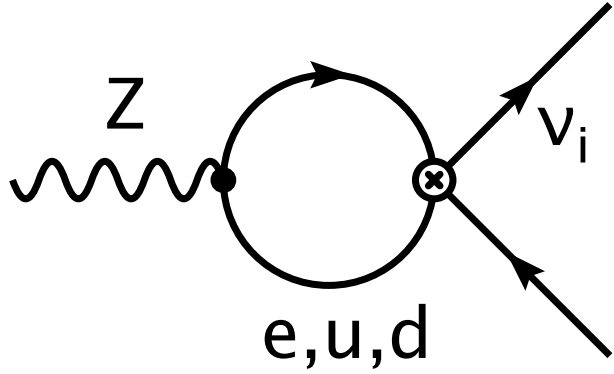}}
\par}
\caption{Contributions of lepton flavor conserving NSI to the Z invisible decay width.\label{fig:Z-inv}}
\end{figure}

\begin{figure}
{\par \centering\resizebox*{.5\textwidth}{!}{\includegraphics{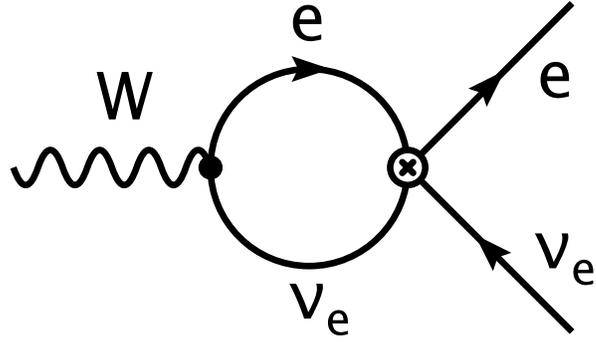}}
\par}
\caption{Contributions of leptonic flavor conserving NSI to the vertex of the W gauge boson.\label{fig:W}}
\end{figure}

For lepton flavor conserving operators, diagrams like
Fig.~\ref{fig:boxEFT} also give rise to interactions with four charged
leptons, however the information we can extract from them is not so
useful. In addition, in that case, there are other interesting
interactions that can also be generated through radiative
corrections. In particular, loop corrections involving interactions
like the ones in eq.~(\ref{eps}) can affect in a non-universal way the
decay rates of the electroweak gauge bosons as shown in Figs.~
\ref{fig:Z-eud}-\ref{fig:W}.  This type of interactions also occur in
the SM but in the SM they are universal, they have the same strength
for the different generations.  Then, we can bound them by using the
various universality tests. These are checked, at most, at the level
of \( 0.1\% \), which is of the order of the SM radiative corrections
(apart from \( m_{t} \) corrections or running \( \alpha \)
corrections) which are order \( \frac{\alpha }{\pi } \).  We expect
the one-loop contributions from NSI to give corrections \(
\frac{\alpha }{\pi }\, \varepsilon \) therefore we expect, from these
processes, to obtain very weak bounds \( \varepsilon <1 \).  Let us
see how these bounds arise.

Fig.~\ref{fig:Z-eud} gives corrections to the hadronic and leptonic
decay widths of the $Z$-boson. The NSI contributions to the amplitudes
of these processes interfere with the tree-level SM amplitudes and the
total correction is proportional to the sum of all neutrino couplings.
Since \( \varepsilon ^{fP}_{ee} \),
\( \varepsilon ^{fP}_{\mu \mu } \)
are bounded by other means we will consider only \( \nu _{\tau } \)
interactions.  The calculation of the diagram gives a correction to
the couplings of of \( e \),\( u \),\( d \) given by (\( f=e,u,d \))
\[
\delta g^{f}_{V}=\frac{\alpha }{12\pi s^{2}_{W}c^{2}_{W}}
\ln \left(\frac{\Lambda}{m_{Z}}\right) \sum _{\alpha=e,\mu ,\tau }
\sum_P\varepsilon ^{fP}_{\alpha\alpha}~,\] 
\[
\delta g^{f}_{A}=\frac{\alpha }{12\pi s^{2}_{W}c^{2}_{W}}
\ln \left( \frac{\Lambda}{m_{Z}}\right) 
\sum _{\alpha=e,\mu ,\tau }
\sum_P (-1)^P \varepsilon ^{fP}_{\alpha\alpha}~,
\]
where we have adopted the PDG convention for SM $Z$-couplings to
fermions,
\[
\mathcal{L}_Z=-\frac{e}{2s_{W}c_{W}}\sum _{f}\, \overline{\psi _{f}}
\gamma ^{\rho }(g^{f}_{V}-g^{f}_{A}\gamma _{5})\psi _{f}Z_{\rho }~,
\]
with $g_V=g_L + g_R$, $g_A = g_L - g_R$ (see table~\ref{tab:gAi}), and
\( (-1)^{P}=+1 \) for \( P=L \) and \( (-1)^{P}=-1 \) for \( P=R \).
Results for \( g^{e}_{V,A} \) are usually presented in terms of an
effective
\( g^{e}_{V,A} \) at the $Z$ peak which incorporates the SM electroweak
radiative corrections (but not final state QED corrections).
\( g^{e}_{A} \) is known with a good precision (\( g^{e}_{A}=-0.50111\pm 0.00035 \)
without assuming universality in the fit) and agrees quite well with
the SM prediction for a light Higgs $g^{e}_A \approx -0.5012$,
therefore additional contributions must be small.  Requiring that \(
\delta g^{e}_{A} \) is smaller, in absolute value, than the error, and
taking $\ln(\Lambda/m_Z) > 1$, we find
\[
\sum_P\sum _{\alpha=e,\mu ,\tau }\varepsilon ^{eP}_{\alpha\alpha}<
\frac{12\pi s^{2}_{W}c^{2}_{W}}{\alpha }\, 0.00035\times 1.64 \approx 0.5\, ,\,
\, \, 90\%\mathrm{CL}~.
\]
For the vector part the precision is not so good and the agreement
with the standard model is not so perfect: the measurement yields
$g^{e}_V \approx -0.03816\pm 0.00047$ while the SM prediction for a
light Higgs is about $g^{e}_V \approx -0.037$. Requiring that the
additional corrections are smaller, in absolute value, than the error
we obtain
\[
\sum_P\sum _{\alpha=e,\mu ,\tau }(-1)^P\varepsilon ^{eP}_{\alpha\alpha}<
\frac{12\pi s^{2}_{W}c^{2}_{W}}{\alpha} \,0.00047\times 1.64 \approx 0.7\, ,\,
\, \, 90\%\mathrm{CL}~.
\]
These bounds are comparable to the tree-level limits one can set from
$e^+ e^- \rightarrow \gamma \nu \bar{\nu}$ and, in particular, if we
consider only couplings to $\nu_\tau$ and assume no cancellations
among different operators we find
\begin{equation}
\big|\varepsilon_{\tau\tau}^{eP}\big|<0.5~.
\end{equation}
If we allow for cancellations among left- and right-handed operators
the limit is slightly moved to about $0.6$.

Although the same corrections appear also for light quarks, they only
affect the total hadronic decay width of the $Z$, which also contains
contributions from $s$,$c$ and $b$ quarks. One can try to subtract the
$b$ and the $c$ contributions but the $s$ quark is practically
impossible to separate. Therefore, although with additional
assumptions (for instance that the $b$, $c$ and $s$ quark couplings
are standard) one can set some bounds on the hadronic $\varepsilon$'s,
the bounds will not be comparable to the ones obtained for the
electron couplings.

On the other hand, Fig.~\ref{fig:Z-inv} gives corrections to the
invisible decay width of the $Z$-boson which, assuming 3 neutrino
flavours, can be used to set bounds on the
\( \varepsilon  \)'s. In this case non-diagonal neutrino NSI could
also contribute. However, these contributions do not interfere with
tree level amplitudes and are already bounded by lepton flavor
violating processes. Neglecting the non-diagonal terms,
Fig.~\ref{fig:Z-inv} will be proportional to the sum of all the \(
\varepsilon \)'s. Evaluation of the diagram yields the following
correction to the $Z$-couplings to neutrinos
\[
\delta g^{\nu _{\alpha}}_{V}=\delta g^{\nu _{\alpha}}_{A}=
\frac{\alpha }{12\pi s^{2}_{W}c^{2}_{W}}
\ln \left( \frac{\Lambda}{m_{Z}}\right) 
\sum _{f=e,u,d}N_{C}(f)\sum_P (g^{f}_{V}+(-1)^{P}g^{f}_{A})\varepsilon
^{fP}_{\alpha\alpha}~.
\]

Assuming that all the other couplings are standard one can obtain
information on the above combination of $\varepsilon$'s from the
invisible decay width of the $Z$, $\Gamma_{\mathrm inv}$.  SM limits
on $\Delta N_\nu$, the number of neutrino species, are in fact bounds
on $\Delta \Gamma_{\mathrm inv}$. If only $\nu_\tau$ NSI contribute to
$\Gamma_{\mathrm inv}$ ($\nu_e$ and $\nu_\mu$ interactions are already
well bounded from other processes) we can set bounds on
$\varepsilon^{fP}_{\tau\tau}$.  Using that the $\Gamma_{\mathrm inv}$
is proportional to $\sum_\alpha (g^{\nu_\alpha }_A)^2$ (we assume only
left-handed neutrinos so $g^{\nu_\alpha }_A$ = $g^{\nu_\alpha }_V$) one
can easily see that
\begin{equation}
\frac {\delta g^{\nu_\tau}_A}{g^{\nu_\tau}_A}= 
\frac{1}{2}\Delta N_\nu~,
\end{equation}
where in the denominator we have assumed approximate universality in
the couplings.  Taking only one operator at a time and requiring that
the NSI contribution is smaller than the error in absolute value
($\Delta N_\nu < 0.008$) we find (at 90\% CL)
\begin{eqnarray}
\left|\varepsilon^{uL}_{\tau\tau}\right| &<&1.4~,\\
\left|\varepsilon^{uR}_{\tau\tau}\right| &<&3~,\\
\left|\varepsilon^{dL}_{\tau\tau}\right| &<&1.1~,\\
\left|\varepsilon^{dR}_{\tau\tau}\right| &<&6~,\\
\left|\varepsilon^{eL}_{\tau\tau}\right| &<&5~,\\
\left|\varepsilon^{eR}_{\tau\tau}\right| &<&6~.
\end{eqnarray}
Of course for electron couplings to neutrinos we have much better
bounds from the leptonic decay widths of the $Z$. For hadronic
couplings these are the best limits we have.

In addition to the processes considered there are other one-loop
diagrams that could give interesting contributions. For instance, the
diagram in Fig.~\ref{fig:W} gives corrections to the coupling of the
W-boson to the electron only. This coupling can be absorbed in the
definition of \( G_{F} \) affecting equally \( \mu \) decay and \(
\beta \) decays (or \( \tau \rightarrow e\, \nu _{\tau }\, \bar{\nu
}_{e} \) and \( \pi \rightarrow e\, \bar{\nu }_{e} \)). Note, however,
that it will not affect \( \tau \rightarrow \mu \, \nu _{\tau }\,
\bar{\nu }_{\mu } \) or
\( \pi \rightarrow \mu \, \bar{\nu }_{\mu } \). Obviously, these contributions
also will affect differently \( W\rightarrow e\, \bar{\nu }_{e} \) and
\( W\rightarrow \mu \, \bar{\nu }_{\mu } \) or \( W\rightarrow \tau \,
\bar{\nu }_{\tau } \). Therefore, universality limits can be used to
set bounds on \( \varepsilon ^{eL}_{ee} \), the only coupling
appearing Fig.~\ref{fig:W}. However, we do not expect interesting
bounds since this coupling is already bounded from $\nu_e e
\rightarrow \nu_e e$ scattering.

\begin{figure}
{\par\centering\resizebox*{.5\textwidth}{!}{\includegraphics{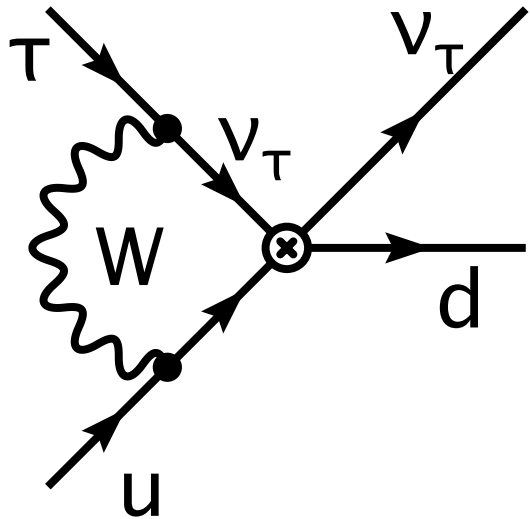}}
\par}
\caption{Contributions of lepton flavor conserving NSI to standard charged
current processes.\label{fig:taubox}}
\end{figure}

Finally one can use the recent determinations of $G_F$ in $\tau$
decays (such as $\tau \rightarrow \nu_\tau \mu \bar{\nu}_\mu$, $\tau
\rightarrow \nu \pi$) to set constraints
\cite{Pich:2001rj} on new physics involving the $\tau$.
Operators of the form $(\bar{\tau} \gamma ^\rho \nu_\tau)(f'
\gamma_\rho P f)$ are induced at one loop by the operators we are
considering (see Fig.~\ref{fig:taubox}). Thus, from the bounds in
\cite{Pich:2001rj} one can extract limits on the flavour diagonal
$\varepsilon^{fL}_{\tau \tau}$.  However, these limits are loser than
the ones already discussed and only affect left-handed couplings.

In any case, one can use atmospheric neutrino data to set much more
stringent limits on these couplings, once we use that the
$\varepsilon^{fP}_{\mu\mu}$ are well bounded from $\nu_\mu$ scattering
experiments.

\section{KamLAND and SNO/SK}
\label{kam-sno}

The large mixing angle (LMA) oscillation solution of the solar
neutrino problem \cite{bahcalldavies}, has been confirmed by the
anti--neutrino reactor experiment KamLAND.  It is a short baseline
experiment (matter effects are negligible in the present global
allowed region by solar and reactor data) measuring electron
anti-neutrinos through a charged current process.  We will assume that
KamLAND finds an oscillatory signal, so the neutrino parameters
$\Delta m^2_{sol}$ and $\theta_{12}$ will be determined with good
precision. Let us stress that the neutral current NSI we are
discussing in this work does not affect in a significant way the
KamLAND observables. On the other hand, the solar neutrino data may be
plagued by NSI present in the matter potential (evolution in the Sun
and in the Earth) and/or by NSI present in the neutrino neutral
current detection.  Therefore the consistency of the KamLAND and the
solar neutrino data can give us information on NSI that contribute in
solar neutrino experiments.  In this section, we anticipate the
constraints that could be set with three years of KamLAND data.

We will consider solar neutrino data only from SNO and SK (for a
 discussion on the impact of NSI in the Borexino observables, see
 Ref.~\cite{Berezhiani:2001rt}).  The main source of neutrinos at the
 energies relevant for SNO and SK is $^8B$ neutrinos. There is also a
 small contribution due to hep neutrinos (roughly 0.5\% of the total
 rate if we use the solar standard model fluxes) that we consider as a
 source of systematic error in the experiments. The reason for using
 only $^8B$ neutrino data is that the resulting bounds on NSI are
 almost solar model independent. For simplicity, we will use the total
 number of events measured at these experiments for the daytime and
 for the nighttime characterized by the total rates and the day--night
 asymmetries.  These data are enough to show the bounds that can be
 reached by this set of experiments. A more complete analysis should
 include spectral SK and SNO information.

NSI modify the evolution in matter by the effective parameters
$\varepsilon'^{V}$ and $\varepsilon^{V}$, described as a function of
the fundamental NSI by
\footnote{In this analysis we have taken $\theta_{13}=0$, which is a 
good approximation because solar and KamLAND data are weakly sensitive to 
$\theta_{13}$ below the CHOOZ bound.} 
\bea
\varepsilon^{V}=\sum_{f=e,u,d}(\cos\theta_{23}\varepsilon_{e \mu}^{f V}
-\sin\theta_{23}\varepsilon_{e \tau}^{f V}) \frac{N_f}{N_e},
\eea
\bea
\varepsilon'^{V}=\sum_{f=e,u,d}(\varepsilon_{e e}^{f V}
-(\cos^2\theta_{23}\varepsilon_{\mu \mu}^{f V}
+\sin^2\theta_{23}\varepsilon_{\tau \tau}^{f V}
-\sin2\theta_{23}\varepsilon_{\mu \tau}^{f V})) \frac{N_f}{N_e},
\label{epsilonp}
\eea
where $N_f$ is the number density of target particles $ f$ in matter,
and $\varepsilon^{V} =\varepsilon^L +
\varepsilon^R $. We are using the fact that the 3-neutrino evolution in 
matter can be described in good approximation by an effective
2-neutrino description even for $\varepsilon_{e \tau} \sim 1$, because
$G_F N_e \ll \Delta m^2_{atm}/E_\nu$.  
The neutrino evolution
Hamiltonian in the presence of NSI can be found in \cite{fitNSI}.

Let us discuss the main effect of NSI in the solar neutrino evolution:
\begin{itemize}
\item
 $\varepsilon'^{V}$ modifies the survival probability through the
change of the effective matter density seen by the neutrinos. If
$\varepsilon'^{V}$ is negative, the effective density is smaller and
the resonance happens at larger energies, changing the recoil energy
spectrum and the total rate in SK and SNO. Thus, the evolution of the
$^8B$ neutrinos in the Sun is sensitive to negative
$\varepsilon'^{V}$. On the contrary, a positive $\varepsilon'^{V}$
increases the effective density and the resonance happens at lower
energies, having a flat spectrum for the $^8B$ neutrinos and a
slightly affected rate. This would imply that there is no bound on
positive $\varepsilon'^{V}$ from $^8B$ neutrinos. However, the large
effective density would be also in the Earth and it would make an
effect in the day--night difference. The measured day-night asymmetry
($A_{DN}=2(N-D)/(N+D)$ where D(N) are the number of events measured
during the daytime (nighttime)) bounds positive $\varepsilon'^{V}$.
\item
$\varepsilon^{V}$, if small, modifies the survival probability through
the change of the effective mixing \cite{lisiNSI}. Roughly speaking,
KamLAND depends on $\theta$ while SNO/SK observables depend on the
combinations $\theta+\varepsilon^{V}$ and the comparison of the
allowed ranges for the mixing angle extracted from KamLAND data and
from the SK--SNO data define the $\varepsilon^{V}$ bounds we can get.
\end{itemize}

We consider the following solar neutrino observables:

SNO Charged Current (CC): NSI appear only in the propagation due to
 matter effects (only vector couplings contribute),
\bea
[CC] &=& f_B \langle
P_{ee}(\varepsilon'^{V},\varepsilon^{V})\rangle_{CC}~,
\label{snocc}
\eea
where
\bea
f_B=\frac{\Phi_{^8B}}{\Phi^{SSM}_{^8B}},
\eea
is the $^8$B solar neutrino flux normalized to the standard solar
model prediction. We denoted [XX] as the observable XX normalized to
the case of no transition ($[XX]=\frac{XX(P_{ee})}{XX(P_{ee}=1)}$) and
$\langle~\rangle_{XX}$ indicates the observable XX averaged with the
detector response.  $P_{ee}$ is the probability that $\nu_e$ produced
in the sun will arrive as $\nu_e$ at the detector.

SNO Neutral Current: NSI appear mainly in the neutrino detection. At
low energies the neutrino--deuteron cross section is dominated by the
Gamow-Teller transitions, so that the cross section scales as $g^2_A$,
where $g_A$ is the coupling of the neutrino current to the axial
isovector hadronic current 
\footnote {We are grateful to Jos\'e Bernab\'eu for an enlightening  
discussion about this point.}
\cite{bahcall,bernabeu,ando,beacomparke,Chen:2002pv}. 
In the SM, $g_A=g^u_A-g^d_A=1$ and using that the nuclear
corrections to $g_A$ are the same when the NSI are added we
obtain:
\bea
[NC] &\sim& f_B (1+2\varepsilon^{A})~,
\label{snonc}
\eea
where
\bea
\varepsilon^{A} \sim \sum_{\alpha=e,\mu,\tau} \langle P_{e \alpha} \rangle_{NC} (\varepsilon_{\alpha \alpha}^{u A}-\varepsilon_{\alpha \alpha}^{d A}),
\eea

Thus, this cross section is mainly sensitive to the axial part of the
NSI and contains complementary information to the oscillation
probabilities that depend on the vector part of the NSI. Notice that
in the case of absence of axial NSI, the NC detection is blind to
oscillations ($\sum_{\alpha=e,\mu,\tau} P_{e\alpha}=1$) and determine
the total $^8$B flux even if neutrinos oscillate \cite{chen}.

SNO--SK Elastic Scattering: NSI appear in the neutrino evolution and
also in the neutrino detection \cite{Berezhiani:2001rt}.
\bea
[ES] &=& f_B
\left[ r_e \langle P_{ee}(\varepsilon'^{V},\varepsilon^{V})\rangle_{ES}
+0.157 r_a (1-\langle
P_{ee}(\varepsilon'^{V},\varepsilon^{V})\rangle_{ES})
\right]~,
\label{skes}
\eea
where
\bea
r_e=\frac{1}{\sigma_{\nu_e e}(g_L,g_R)}\sigma_{\nu_e
e}(g_L+\varepsilon^{eL}_{ee},g_R+
\varepsilon^{eR}_{ee},\sum_{\alpha \neq e}|\varepsilon^{eP}_{\alpha e}|^2),
\eea
\bea
r_a&=&\frac{\cos^2\theta_{23}}{\sigma_{\nu_\mu
e}(g_L,g_R)}\sigma_{\nu_\mu
e}(g_L+\varepsilon^{eL}_{\mu\mu},g_R+\varepsilon^{eR}_{\mu\mu},\sum_{\alpha
\neq \mu} |\varepsilon^{eP}_{\alpha \mu}|^2)\nonumber\\
&+&\frac{\sin^2\theta_{23}}{\sigma_{\nu_\tau
e}(g_L,g_R)}\sigma_{\nu_\tau
e}(g_L+\varepsilon^{eL}_{\tau\tau},g_R+\varepsilon^{eR}_{\tau\tau},\sum_{\alpha
\neq \tau} |\varepsilon^{eP}_{\alpha \tau}|^2).
\eea
$\sigma_{\nu_e e}(g_L+\varepsilon^{eL}_{ee}...)$ is given in eq.
(\ref{enue}); the $\sigma_{\nu_\mu
e}(g_L+\varepsilon^{eL}_{\mu\mu}...)$ and $\sigma_{\nu_\tau
e}(g_L+\varepsilon^{eL}_{\tau\tau}...)$ are the obvious modifications
of this equation.  In the following, we refer to [ES] as the averaged
value from SK and SNO electron scattering measurements.

Regarding the data used in the present analysis, we assume that in the
next three years:\

- KamLAND will find a clear signal of oscillations (total rate and
energy distortion). From KamLAND, we will be able to know the ranges
of the solar oscillation parameters, $\Delta m^2_{sol}$ and
$\theta_{12}$ with good precision.  Assuming that KamLAND confirm the
present best fit point to the solar neutrino data, we expect $\langle
P_{ee}(\varepsilon'^{V}=0,\varepsilon^{V}=0)\rangle=0.32 (1\pm
0.07)$~\cite{sterile}.

- We assume a moderate improvement on the SNO and SK measurements and
that the NC is measured independently from the CC in the second and
the third SNO phases \cite{snophases}.  The central values correspond
to the present central measurements : [CC]= 0.35 (1$\pm$ 0.05), [NC]=
1.01 (1$\pm$ 0.08) and [ES]= 0.47 (1$\pm$ 0.03) (the present
measurements and errors are [CC]= 0.349 (1$\pm$ 0.057) \cite{sno},
[NC]= 1.008 (1$\pm$ 0.125) \cite{sno}, [ES]= 0.465 (1$\pm$ 0.032)
\cite{sk}).

- We use the day-night asymmetry measured at Super-Kamiokande
$A_{DN}=0.021\pm 0.024$ \cite{sk}, consistent with the first SNO
measurement $A_{DN}=0.07\pm 0.05$ \cite{snodn}.

- In the calculations, we use $0.4\le \cos^2\theta_{23} \le 0.6$, the
expected 90\% CL allowed range from MINOS \cite{minos}.

Finally, the results presented in this section have been obtained
solving numerically the evolution in the Sun and in the Earth using
the electron and neutron number densities from the standard solar
model \cite{bp00} and the preliminary reference earth model
\cite{prem}. Details on the KamLAND simulation used in this analysis
can be found in Ref.~\cite{sterile}. CC and NC cross sections were
obtained from Ref.~\cite{ando}. ES was computed including radiative
corrections \cite{cses}.

We present bounds on a given four-fermion operator, assuming that all
other four-fermion operators are zero.  Let us illustrate how the
bounds come out for a particular NSI, namely $(\bar{e} \gamma^{\rho} L
{e})(\bar{\nu}_{e} \gamma_{\rho} L \nu_e )$.  If we consider only this
operator, eqs.~(\ref{snocc})--(\ref{skes}) simplify to:
\bea
[CC] &=& f_B \langle P_{ee}(\varepsilon^{eL}_{ee})\rangle_{CC}~,
\label{eq1}
\eea
\bea
[NC] &=& f_B ~,
\label{eq2}
\eea
\bea
[ES] &=& f_B (r_e \langle P_{ee}(\varepsilon^{eL}_{ee})\rangle_{ES}
+0.157 (1-\langle P_{ee}(\varepsilon^{eL}_{ee})\rangle_{ES}))~.
\label{eq3}
\eea
In that case, the NC value is a measurement of the $^8$B flux, $f_B$=
1.01 (1 $\pm$ 0.08). This result for $f_B$ combined with eq.~(\ref{eq1})
gives $\langle P_{ee}(\varepsilon^{eL}_{ee})\rangle_{CC}=0.33 (1 \pm
0.09)$.  By inspection, if we compare this averaged rate with the
result from the KamLAND allowed region, $\langle
P_{ee}(\varepsilon'^{V}=0,\varepsilon^{V}=0)\rangle=0.32 (1\pm 0.07)$,
we can conclude that $\varepsilon^{eL}_{ee}$ must be constrained in
the range where matter effects from the NSI are small.  Using this
conclusion in eq.~(\ref{eq3}), $\varepsilon^{eL}_{ee}$ is further
constrained by the ES dependence on $r_e$. Solving numerically the set
of eqs.~(\ref{eq1})--(\ref{eq3}) as a function of $f_B$ and
$\varepsilon^{eL}_{ee}$ and using the allowed range of parameters
($\Delta m^2_{sol}$,$\theta_{12}$) from KamLAND, we get the 90\%CL
bound :
\bea
-0.08 < \varepsilon^{eL}_{ee} < 0.10~.
\eea

More precisely, given the solar observables considered, we computed
$\chi^2_{sol}$ as a function of the NSI parameter
$\varepsilon^{fP}_{\alpha\beta}$, the oscillation parameters $\Delta
m^2_{sol}$, $\theta_{12}$ and $\theta_{23}$, and the $^8$B flux
normalization, $f_B$.  We add the $\chi^2$ of the KamLAND analysis, a
function of $\Delta m^2_{sol}$ and $\theta_{12}$, to the
$\chi^2_{sol}$. Next, we find the marginalized
$\chi^2(\varepsilon^{fP}_{\alpha\beta})$ by minimizing the total
$\chi^2$, $\chi^2_{total}=\chi^2_{sol}+\chi^2_{KamLAND}$, respect to
the oscillation parameters and the $^8$B flux normalization for each
$\varepsilon^{fP}_{\alpha\beta}$.  Finally, we get the bounds on the
NSI parameter by comparing the function $\Delta
\chi^2(\varepsilon^{fP}_{\alpha\beta})=
\chi^2(\varepsilon^{fP}_{\alpha\beta})-\chi^2_{min}$ and the statistical 
$\chi^2$ distribution with 1 dof ($\chi^2_{min}$ is the minimum of 
$\chi^2$(total) in the full space of parameters). For the different operators 
(one by one) that appear in the solar observables, we get at 90\% CL:
\bea
-0.2 < \varepsilon^{eL}_{\alpha\alpha} < 0.3 \\ 
-0.3 < \varepsilon^{eR}_{ee} < 0.5 \\ 
-0.9 < \varepsilon^{eR}_{\alpha\alpha} < 0.3 \\ 
-0.25 < \varepsilon^{uL}_{ee} < 0.2 \\ 
-0.3 < \varepsilon^{uL}_{\alpha\alpha} < 0.25 \\ 
-0.2 < \varepsilon^{uR}_{ee} < 0.25 \\ 
-0.25 < \varepsilon^{uR}_{\alpha\alpha} < 0.3 \\ 
-0.2 < \varepsilon^{dL}_{ee} < 0.25 \\ 
-0.25 < \varepsilon^{dL}_{\alpha\alpha} < 0.3 \\ 
-0.25 < \varepsilon^{dR}_{ee} < 0.2 \\ 
-0.3 < \varepsilon^{dR}_{\alpha\alpha} < 0.25
\eea
where $\alpha = \mu,\tau$ and
\bea
-0.2 < \varepsilon^{fP}_{\alpha\beta} < 0.3
\eea
where $f=e,u,d$, $P=L,R$, and $\alpha,\beta=e,\mu,\tau$ with $\alpha
\ne \beta$.

\section{$\nu$Factory}
\label{nufact}

In this section, we estimate the sensitivity to NSI of $\nu e$
scattering and $\nu$-DIS at the near detector of a neutrino
factory. We outline in the next paragraphs the order of magnitude of
the limits the near detector could set on $\varepsilon_{\alpha
\alpha}$ ($ \alpha \neq
\tau$), and  $\varepsilon_{\alpha \beta}$ ($ \alpha \neq
\beta$).  A more careful analysis follows
in subsections \ref{5.1} and \ref{5.2}, based on the $\nu$-DIS chapter
of the ECFA-CERN Neutrino Factory Study \cite{Mangano:2001mj}.
Mangano {\it et al.} \cite{Mangano:2001mj} discuss various
measurements from which $\sin^2 \theta_W$ could be extracted, with
their potential errors.  Finally, in section \ref{5.3}, we briefly
review potential nufactory bounds on new charged current neutrino
interactions \cite{Huber:2001de,Mangano:2001mj}.

  For flavour diagonal NSI involving $\nu_e$ and $\nu_\mu$, $\sw2$
measurements at a near detector of a neutrino factory should be more
sensitive than oscillation probabilities measured at the far detector.
$\varepsilon_{\alpha \alpha }$ will interfere with the SM amplitude
for $\nu_\alpha f \rightarrow \nu_\alpha f$,
\footnote{$\alpha \neq \tau$ here} so the diagonal
NSI contribute linearly to scattering.  They also contribute linearly
to the oscillation probability at the far detector, via their MSW
contribution to the neutrino mass matrix. But there should be more
events at the near detector, because the near beam is narrow, so a
well-instrumented detector, of area larger than the beam, can be built
at reasonable cost. This suggests that the statistical and systematic
errors on $\varepsilon_{ee}$ and $\varepsilon_{ \mu \mu}$ would be
smaller at a near detector.

The situation is less clear for flavour {\it changing} $\varepsilon_{
\alpha \beta} (\alpha \neq \beta) $ which contribute quadratically to
$\sw2$.  Neglecting for the moment systematic errors, one can estimate
that the near detector would be sensitive to NSI such that $N^n_{NSI}
\gsim \sqrt{N^n_{SM}}$.  The number of events in the near detector due
to NSI is $N^n_{NSI} \propto \varepsilon^2 N^n_{SM} $, where
$N^n_{SM}$ is the number of SM $\nu$ scattering events.  In the
analysis of \cite{Mangano:2001mj}, $N^n_{SM} \sim 10^{8}$ was taken,
so $\varepsilon_{\alpha \beta} \gsim .01$ might be seen.  At the far
detector \cite{Johnson:1999ci,Ota:2001pw,Huber:2001zw}, 
$\varepsilon_{\alpha
\beta}$s ( with $ \alpha \neq \beta$) appear linearly in the
oscillation probability, because they interfere with the SM
oscillation amplitude.  NSI could disrupt the measurement of
oscillation parameters at the far detector if $N^f_{NSI} \gsim
N^f_{osc}$, where $ N^f_{osc}$ is the number of events at the far
detector due to oscillations.  So $\varepsilon_{\alpha \beta} \gsim
\sin \theta_{13}$ could   interfere with the measurement
of $\sin \theta_{13}$ at the far detector \cite{Huber:2001de}.  A
recent analysis \cite{Freund:2001ui} suggests that a neutrino factory
can realistically measure $\sin \theta_{13} \gsim .01$.  We find that
the near detector would be sensitive to $\varepsilon_{\alpha \beta}
\gsim$ a few $\times .01$, so the sensitivities of the near and far
detector to flavour-changing NSI are similar.

\subsection{Measuring $\sin^2 \theta_W$ leptonically}
\label{5.1}

The weak mixing angle could be measured leptonically in the scattering
 of neutrinos off electrons in the target.  The error is smaller using
 the $\bar{\nu}_\mu, \nu_e$ from the $\mu^+$ beam; $\sin^2 \theta_W$
 can be determined from
\beq
\sigma_{\nu fact} = \sigma(\bar{\nu}_{\mu} e \rightarrow \bar{\nu} e)
+ \sigma ({\nu}_{e} e \rightarrow {\nu} e)
\eeq
with a statistical error of order $\Delta \sin^2 \theta_W = 2 \times
 10^{-4}$ \cite{Mangano:2001mj}.  To estimate ``90 $\%$ C.L.'' limits
 on NSI, we multiply by 1.6, so use $\Delta \sin^2 \theta_W = 3 \times
 10^{-4}$.  The second of these two cross-sections, augmented by the
 contributions of NSI, is given in eq. (\ref{enue}), and the first is
\beq
\sigma (\bar{\nu}_{\mu} e \rightarrow \bar{\nu} e)  =
\frac{2 G_F^2 m_e E_\nu}{\pi}\left[
 \frac{1}{3}(g^e_L + \varepsilon^{eL}_{\mu \mu} )^2 + \frac{1}{3}
\sum_{\alpha \neq \mu} | \varepsilon^{eL}_{\mu \alpha} |^2 +(g_R^e +
\varepsilon^{eR}_{\mu \mu})^2 + \sum_{\alpha \neq \mu} |
\varepsilon^{eR}_{\mu \alpha} |^2 \right]
\label{SM1}
\eeq
Requiring $(\sigma_{\nu fact} - \sigma_{SM})
\lsim  \Delta s_W^2 \times \partial \sigma_{SM}/\partial s_W^2$ gives
\bea
 \frac{2}{3} g^e_L \varepsilon^{eL}_{\mu \mu}+ \frac{1}{3}
 \sum_{\alpha} | \varepsilon^{eL}_{\mu \alpha} |^2 + 2 g_R^e
 \varepsilon^{eR}_{\mu \mu} + \sum_{\alpha } | \varepsilon^{eR}_{\mu
 \alpha} |^2 ~~~~~~~~~~~~ && \nonumber\\ + 2(1+g^e_L)
 \varepsilon^{eL}_{ee}+ \sum_{\alpha } | \varepsilon^{eL}_{\alpha e}
 |^2 +\frac{2}{3}g_R^e \varepsilon^{eR}_{ee} + \frac{1}{3}
 \sum_{\alpha } | \varepsilon^{eR}_{\alpha e} |^2 & <& 6 \times
 10^{-4}~~~,
\label{nufactsum}
\eea
so the flavour-changing NSI satisfy
\bea
 |\varepsilon^{eL}_{\mu \tau}| < .04 && |\varepsilon^{eR}_{\mu \tau}|
< .02 \nonumber\\ |\varepsilon^{eL}_{\tau e}| < .02 &&
|\varepsilon^{eR}_{\tau e}| < .04~~~.
\eea
(We neglect $\varepsilon^{eP}_{\mu e}$ because it is more strongly
constrained by $\mu \rightarrow 3 e$---see section \ref{oneloop}.)
The flavour diagonal $\varepsilon$s satisfy
\beq
 |\varepsilon^{eL}_{ \mu \mu}| < 0.003 ~~~, |\varepsilon^{eR}_{\mu
 \mu}| < 0.001 ~~~, |\varepsilon^{eL}_{e e}| < 0.0004 ~~~,
 |\varepsilon^{eR}_{e e}| < 0.004
\eeq
assuming no cancellations in equation (\ref{nufactsum}).

\subsection{Measuring $\sin^2 \theta_W$ in DIS}
\label{5.2}

In neutrino Deep Inelastic Scattering (DIS), the NC events due to
incident $\nu_e$ and $\bar{\nu}_\mu$ cannot be separated.  So the
Paschos-Wolfenstein ratio:
\beq
R_{PW} = \frac{ \sigma_{NC} (\nu_\mu) - \sigma_{NC} (\bar{\nu}_\mu)} {
\sigma_{CC} (\nu_\mu) - \sigma_{CC} (\bar{\nu}_\mu)}
\eeq
 is not available, and $s_W^2$ cannot be so elegantly disentangled
from parton distributions.  It was conservatively estimated in
\cite{Mangano:2001mj} that a neutrino factory could measure $s_W^2$ to
one part in $10^3$, via a ratio of the form
\bea
R_\mu^- & = &\frac{ \sigma_{NC} (\nu_\mu) + \sigma_{NC} (\bar{\nu}_e)}
{ \sigma_{CC} (\nu_\mu) + \sigma_{CC} (\bar{\nu}_e)} \nonumber \\ & =
& \frac{(\tilde{g}_{L \mu} )^2 + (\tilde{g}_{R e})^2 + r
[(\tilde{g}_{R \mu})^2 + (\tilde{g}_{ L e})^2]} {1+r}
\eea
where $r \sim 0.5$ is defined in eq.(\ref{eqnr}), and the $\tilde{g}$
with lepton subscripts are as defined in equations
(\ref{gl2}),(\ref{gr2}), replacing $e \rightarrow \mu$ as required.

A similar ratio $R^+$ can be defined for the beam produced in the
decay of a $\mu^+$:
\bea
R_\mu^+ & = &\frac{ \sigma_{NC} (\bar{\nu}_\mu) + \sigma_{NC} (\nu_e)}
{ \sigma_{CC} (\bar{\nu}_\mu) + \sigma_{CC} (\nu_e)} \nonumber \\ & =
& \frac{(\tilde{g}_{R\mu})^2 + (\tilde{g}_{L e})^2 + r [(\tilde{g}_{L
\mu})^2 + (\tilde{g}_{R e})^2]} {1+r}
\eea
$R^+$ and $R^-$ can be measured with equal sensitivity, so $s_W^2$ can
 be independently determined from both. The NSI appear in one or the
 other not multiplied by $r$, so we can set bounds of order $\Delta
 (\tilde{g}_{P \mu,e})^2
\lsim \partial R^{\pm}/ \partial s_W^2 \times 10^{-3} \sim .0005$, for instance
\bea
\frac{2 g_L^u \varepsilon^{uL}_{\mu \mu}}{1+r} & < & 5 \times 10^{-4}
\eea
or equivalently
\beq
|\varepsilon^{uL}_{\mu \mu}| < 1 \times 10^{-3} ~~~,
\eeq
and assuming no cancellations among the terms in $\tilde{g}_{P \ell}$
(eqs.(\ref{gl2}),(\ref{gr2})) :
\bea
|\varepsilon^{uL}_{\mu \mu}|, |\varepsilon^{uL}_{ee}| < 1 \times
 10^{-3}, && |\varepsilon^{dL}_{\mu \mu}|,|\varepsilon^{dL}_{ee}| < 9
 \times 10^{-4} \nonumber \\ |\varepsilon^{uR}_{\mu \mu}|,
 |\varepsilon^{uR}_{ee}| < 2 \times 10^{-3}, && |\varepsilon^{dR}_{\mu
 \mu}|,|\varepsilon^{dR}_{ee}| < 5 \times 10^{-3} \nonumber \\
 |\varepsilon^{qL}_{\tau \mu}| ,|\varepsilon^{qL}_{\tau e}|,
 |\varepsilon^{qR}_{\tau \mu}| ,|\varepsilon^{qR}_{\tau e}| & < & 3
 \times 10^{-2} ~~~.
\eea

\subsection{Charged current interactions}
\label{5.3}

Non standard neutrino interactions of the form given in equation
(\ref{eps}) could contribute at one loop to muon decay or charged
current $\nu$ scattering off quarks in the near detector.  This would
occur via the external dressing of the four fermion operator with a
$W$ loop, as discussed in section
\ref{oneloop}. For instance, exchanging a $W$ between the $\nu_\mu$ and
$e$ legs of $\varepsilon (\bar{\nu}_\tau \gamma^\rho \nu_\mu)
(\bar{e}\gamma_\rho L e)$ would generate the operator $c \varepsilon
(\bar{\nu}_\tau \gamma^\rho \nu_e) (\bar{e}\gamma_\rho L \mu)$. This
would produce a $\nu_\tau$ from $\mu$ decay, which could turn into a
$\tau$ in CC scattering off quarks. The sensitivity of the near
detector to operators of the form $ 2 \sqrt{2} G_F \eta
(\bar{e}\gamma_\rho \mu) (\bar{\nu}_\beta \gamma^\rho \nu_\alpha)$ was
estimated in \cite{Mangano:2001mj} (see also
\cite{Datta:2000ci,Huber:2001de,Bueno:2000jy}) to be
\beq
\eta \lsim  10^{-4} \sqrt{N} \times 
\sqrt{\frac{ \rho_{det}}{100g/cm^2}}
\sqrt{\frac{{\rm ~number ~of ~\mu~ decays}}{10^{20}}}
\eeq
 where $N$ is the number of $\tau$ events required in the detector for
a signal. $N \sim 10$ was taken in \cite{Mangano:2001mj}.  For $c \sim
.002$, this is a weaker bound on $\varepsilon^{eP}_{e \tau},
\varepsilon^{eP}_{\mu \tau}$ than from measuring $\sw2$. However, in
the opposite limit of $c \sim 1$ (for instance if the NSI are induced
by dimension 6 operators), it is clear that flavour changing NSI are
more readily detected via CC interactions than NC interactions.

\section{Summary}
\label{summary}

We have considered non-standard interactions of neutrinos with first
generation leptons and quarks, parametrised as
\beq
{\cal L}_{eff}^{NSI} =- \varepsilon^{fP}_{\alpha \beta} 2 \sqrt{2} G_F
 (\bar{\nu}_\alpha \gamma_{\rho} L \nu_\beta) (\bar{f} \gamma^{\rho}P
 f) \ .
\label{epsinsum}
\eeq
We have taken a phenomenological approach, assuming that the new
physics which induces these non-standard neutral current operators
does not generate the $SU(2)$ related charged lepton operators at tree
level.  This could be the case if the operators in (\ref{epsinsum})
are of dimension eight or larger. Then the tight bounds on charged
lepton NSI do not apply, so we have assembled present and future
constraints on such operators from purely neutrino processes.

We point out though, that even if only neutrino neutral current NSI
are present at tree level, they will necessarily induce the related
vertices with charged leptons at one loop, via $W$ exchange.
Moreover, radiative corrections involving these neutrino NSI could
affect a variety of precision observables.  Thus we have also set
bounds on the strength of the NSI in (\ref{epsinsum}) from their one
loop effects, which in some cases are more stringent than the tree
level ones.

Our results are summarized in tables \ref{tab:ffopt} and
\ref{tab:fcffv}.  We list the limit on $\varepsilon^{fP}_{\alpha
\beta}$ that an experiment would set if only one NSI operator was
present.  The limits that arise if cancellations are allowed among the
operators are presented in the body of the paper.

Tree level bounds arise from low energy neutrino scattering
experiments.  We have collected present constraints, including the
recent NuTeV data, and estimated future limits attainable at the near
detector of a neutrino factory from measuring $\sin^2 \theta_W$ (both,
leptonically and in neutrino DIS).  We found that these experiments
are more sensitive to $\varepsilon^{fP}_{ee}$ and
$\varepsilon^{fP}_{\mu \mu}$ than the $\nu$factory far detector.  They
also provide the best bounds on flavour changing interactions
involving $\nu_\tau$.

One loop bounds on lepton flavour violating operators are very
stringent for $\varepsilon^{fP}_{e \mu}$. The present experimental
limits are so strong, that even with the loop suppression, the
$\varepsilon^{fP}_{e \mu}$ are constrained to be ${\cal
O}(10^{-3})$. The analogous bounds on $\varepsilon^{fP}_{\tau \alpha}$
($\alpha=e,\mu$) from several tau decays are just order one, although
the experimental limits on some of these decays could improve, in
which case the relevant bound can be rescaled.  Regarding flavour
diagonal NSI, one loop bounds on $\varepsilon^{eP}_{\tau \tau}$ from
the precise measurement of $g_A^e$ at LEP are of the same order as the
tree level limits from $e^+ e^- \rightarrow \gamma \nu \bar\nu$, while
the limits on $\varepsilon^{qP}_{\tau \tau}$ from the invisible $Z$
width are the only constraints from laboratory data.

Finally, we have estimated future bounds from comparing KamLAND and
solar neutrino data.  These could constrain $\varepsilon^{qP}_{\tau
\tau}$ at the level of ${\cal O}(0.3)$ (which is roughly one order of
magnitude improvement of the present limits), and set bounds on
$\varepsilon^{eP}_{\tau \tau}$ which are comparable to those from LEP.

We hoped to show that long-baseline experiments at a neutrino factory
would indeed measure oscillation parameters, since NSI would be seen
first in other experiments, like short baseline, high intensity,
precision neutrino scattering experiments.  To what degree have we
succeeded?

Suppose that no evidence for NSI is found; does this mean
such interactions can be ignored in long baseline oscillation
experiments?  That is, if the near detector of a neutrino factory sees
no NSI, will the far detector measure oscillation parameters?  The
answer we find seems to be that the NSI cannot quite be ignored. The
three problematic $\varepsilon$s are $\varepsilon_{\tau \tau}$,
$\varepsilon_{\tau \mu}$ and $\varepsilon_{\tau e}$.

The Mangano et al. \cite{Mangano:2001mj} analysis suggests that the
 near detector would be sensitive to $\varepsilon_{\tau \mu},
 \varepsilon_{\tau e} \gsim .02$, and the Freund et al. analysis
 \cite{Freund:2001ui} says the far detector could see $\sin
 \theta_{13} \lsim .01$.  We would like $\varepsilon_{\alpha \beta}<
 \sin \theta_{13}$ to be sure that the $\varepsilon$s do not confuse
 the determination of $\sin \theta_{13}$.  Can these two analyses be
 compared? Freund et al. use more events, but the error in the near
 detector analysis of \cite{Mangano:2001mj} is systematic, so it is
 not clear how the errors scale if we increase the number of events.
The error on the DIS determination of $\sw2$ could be significantly
smaller than estimated in \cite{Mangano:2001mj}, because the source of
error is the parton distribution functions, and the near detector will
measure these \cite{paolo}.  So the near detector might indeed be able
to constrain $\varepsilon^{qP}_{\tau \mu},\varepsilon^{qP}_{\tau e}<
.01$.  Also, $\varepsilon^q_{\tau \mu} \gsim .01$ could possibly be
seen at ICARUS/OPERA \cite{Ota:2002na}, where the neutrino beam is
largely $\nu_\mu$ produced in pion decay, and the detector should be
able to identify $\tau$s.

$\varepsilon^{eP}_{\tau \mu}, \varepsilon^{eP}_{\tau e}$ remain a
problem, but could perhaps be disentangled from $\theta_{13}$ at a
neutrino factory by using the beam spectrum, as discussed in
\cite{CampanelliRomanino}.

\vskip 0.15in

{\bf Acknowledgments}
\vskip 0.15in
We thank Anna Rossi for stimulating our interest in this physics.
We also thank Jose Bernab\'eu, Paolo Gambino, Michele Maltoni and 
Antonio Pich for discussions, and Cecilia Lunardini, Alexander Friedland and
Jose Valle for comments on the manuscript.  This work was partially
supported by the Spanish MCyT grants BFM2002-00345, BFM2002-00568 and
FPA2001-3031, by the TMR network contract HPRN-CT-2000-00148 of the
European Union and by the Generalitat Valenciana grants CTIDIB/2002/24
and GV01-94.  CPG acknowledges support from NSF grant No. PHY-0070928.

\renewcommand{\arraystretch}{1.5}
\begin{table}
\caption{ \label{tab:ffopt} \sf
Flavour conserving four fermion vertices involving two neutrinos and
two first generation fermions ($\bar{e}e$,$\bar{d}d$ or $\bar{u}u$),
the best current and the best future 90 $\%$ CL limits that can be set
on the coefficients $2 \sqrt{2} G_F \varepsilon$ of the four fermion
vertices.  See eq. (\ref{eps}) for the definition of $\varepsilon$.
The limits from processes marked with an asterisk, $\as$, arise at one
loop and are inversely proportional to $\log(\Lambda/m_W)$. We have
assumed $\log(\Lambda/m_W) > 1$ (see section \ref{oneloop}).  }
\begin{tabular}{||  c | c | c ||} \hline \hline
  vertex& current limits& future limit\\
\hline \hline
%
%
 $(\bar{e} \gamma^{\rho} P {e} ) (\bar{\nu}_{\tau} \gamma_{\rho} L
   \nu_\tau )$ & $|\varepsilon^{eP}_{ {\tau} {\tau} }|< 0.5$ & $ -0.2
   < \varepsilon^{eL}_{\tau\tau} < 0.3$ \\ & & $-0.9 <
   \varepsilon^{eR}_{\tau\tau} < 0.3$ \\ & $(g_A^e \, \, @ \, \,{\rm
   LEP})\as$ & KamLAND and SNO/SK \\ \hline
 $(\bar{u} \gamma^{\rho} P {u} ) (\bar{\nu}_{\tau} \gamma_{\rho} L
   \nu_\tau ) $ & $|\varepsilon^{uL}_{ {\tau}\tau }| < 1.4 $ & $-0.3 <
   \varepsilon^{uL}_{\tau\tau} < 0.25$
\\
 & $|\varepsilon^{uR}_{ {\tau}\tau }| < 3 $ & $-0.25 <
\varepsilon^{uR}_{\tau\tau} < 0.3$
\\
 & $(\Gamma_{inv})\as$ & KamLAND and SNO/SK \\ \hline
 $(\bar{d} \gamma^{\rho} L {d} ) (\bar{\nu}_{\tau} \gamma_{\rho} L
   \nu_\tau )$ & $|\varepsilon^{dL}_{ \tau\tau}| < 1.1$ & $-0.25 <
   \varepsilon^{dL}_{\tau\tau} < 0.3$
\\
  & $|\varepsilon^{dR}_{ \tau\tau}| < 6$ & $-0.3 <
  \varepsilon^{dR}_{\tau\tau} < 0.25$
\\
  & $(\Gamma_{inv})\as$ & KamLAND and SNO/SK \\ \hline
%
%
 $(\bar{e} \gamma^{\rho} P {e} ) (\bar{\nu}_{\mu} \gamma_{\rho} L
   \nu_\mu )$ & $|\varepsilon^{e P}_{ \mu \mu}| < 0.03 $ &
   $|\varepsilon^{e L}_{ \mu \mu}| < 0.003 $ \\ & & $
   |\varepsilon^{eR}_{ {\mu} \mu }| < 0.001 $ \\ & CHARM II & leptonic
   $s_W^2$ at nufact \\ \hline
 $(\bar{u} \gamma^{\rho} P {u} ) (\bar{\nu}_{\mu} \gamma_{\rho} L
   \nu_\mu )$ & $|\varepsilon^{uL}_{ {\mu} \mu }| < 0.003 $
   & $|\varepsilon^{uL}_{ {\mu} \mu }| < 0.001$ \\ & $ -0.008 <
   \varepsilon^{uR}_{{\mu} \mu } < 0.003 $ & $|\varepsilon^{uR}_{
   {\mu} \mu }| < 0.002 $ \\ & NuTeV &$s_W^2$ in DIS at nufact \\
   \hline
 $(\bar{d} \gamma^{\rho} P {d} ) (\bar{\nu}_{\mu} \gamma_{\rho} L
   \nu_\mu )$ & $|\varepsilon^{dL}_{ {\mu} \mu}| < 0.003 $ &
   $|\varepsilon^{dL}_{{\mu} \mu }| < 0.0009$ \\ & $-0.008 < 
   \varepsilon^{dR}_{ {\mu} \mu } < 0.015 $ & $|\varepsilon^{dR}_{
   {\mu} \mu }| < 0.005 $\\ & NuTeV &$s_W^2$ in DIS at nufact \\
   \hline
%
%
 $(\bar{e} \gamma^{\rho} P {e} ) (\bar{\nu}_{e} \gamma_{\rho} L \nu_e
   )$ & $-0.07< \varepsilon^{eL}_{ {e} e} <0.1 $ &
   $|\varepsilon^{eL}_{ee} | < 0.0004 $\\ & $-1 < \varepsilon^{eR}_{
   ee} <0.5 $ & $|\varepsilon^{eR}_{ e e} | < 0.004 $\\ & LSND &
   leptonic $s_W^2$ at nufact \\\hline
 $(\bar{u} \gamma^{\rho} P {u} ) (\bar{\nu}_{e} \gamma_{\rho} L \nu_e
   )$ & $-1 < \varepsilon^{uL}_{ ee } < 0.3$ & $|\varepsilon^{uL}_{ ee
   }| < 0.001$ \\ & $ -0.4 < \varepsilon^{uR}_{ ee } < 0.7 $
   &$|\varepsilon^{uR}_{ ee }| < 0.002 $ \\ & CHARM &$s_W^2$ in DIS at
   nufact \\ \hline
 $(\bar{d} \gamma^{\rho} P {d} ) (\bar{\nu}_{e} \gamma_{\rho} L \nu_e
   )$ & $ -0.3 < \varepsilon^{dL}_{ ee} < 0.3$ & $|\varepsilon^{dL}_{
   ee} | < 0.0009$ \\ & $ -0.6 < \varepsilon^{dR}_{ee} < 0.5$ &
   $|\varepsilon^{dR}_{ee} | < 0.005$\\ & CHARM &$s_W^2$ in DIS at
   nufact \\
   \hline \hline
\end{tabular}
\end{table}
\renewcommand{\arraystretch}{1}

\renewcommand{\arraystretch}{1.5}
\begin{table}
\caption{ \label{tab:fcffv} \sf
Flavour changing four fermion vertices involving two neutrinos and two
first generation fermions ($\bar{e}e$,$\bar{d}d$ or $\bar{u}u$), the
best current, and the best future 90 $\%$ CL limits that can be set on
their coefficients $2 \sqrt{2} G_F \varepsilon$.  See eq. (\ref{eps})
for the definition of $\varepsilon$.  The limits from processes marked
with an asterisk, $\as$, arise at one loop and are inversely
proportional to $\log(\Lambda/m_W)$. We have assumed
$\log(\Lambda/m_W) > 1$ (see section \ref{oneloop}).  }
\begin{tabular}{||  c | c | c ||} \hline \hline
  vertex& current limits& future limit\\
\hline \hline
 $(\bar{e} \gamma^{\rho} P {e} )
   (\bar{\nu}_{\tau} \gamma_{\rho} L \nu_\mu )$  &
 $| \varepsilon^{eP}_{\tau
  \mu}| < 1.2$ &
  \\
 &$( \tau \rightarrow  \mu \bar{e} e )\as$ &  
 \\
 & $|  \varepsilon^{eP}_{ \tau
  \mu}| < 0.1$  &  $|  \varepsilon^{eL}_{ \tau
  \mu}| < 0.04 , |\varepsilon^{eR}_{ \tau
 \mu }| < 0.02  $ \\
 &  CHARM II& leptonic $s_W^2$ at nufact \\ \hline
 $(\bar{u} \gamma^{\rho} P {u} )
   (\bar{\nu}_{\tau} \gamma_{\rho} L \nu_\mu )$  &
   $|\varepsilon^{uP}_{ {\tau}
  \mu}| < 2.8$& 
\\
  &$( \tau \rightarrow  \mu \rho)\as$ & 
\\
  & $|  \varepsilon^{uP}_{{\tau}
  \mu}| < 0.05   $ &  $|  \varepsilon^{uP}_{{\tau}
  \mu}| < 0.03   $ \\
 & NuTeV & $s_W^2$ in DIS at nufact \\ \hline
 $(\bar{d} \gamma^{\rho}  P{d} )
   (\bar{\nu}_{\tau} \gamma_{\rho} L \nu_\mu )$ &
 $|  \varepsilon^{dP}_{ {\tau}
  \mu}| <  2.8$&
  \\
  &$( \tau \rightarrow  \mu \rho)\as$ &
  \\
  & $|  \varepsilon^{dP}_{{\tau}
  \mu}| < 0.05   $ &  $|  \varepsilon^{dP}_{{\tau}
  \mu}| < 0.03   $ \\
 & NuTeV & $s_W^2$ in DIS at nufact \\ \hline
%
%
 $(\bar{e} \gamma^{\rho} P {e} )
   (\bar{\nu}_{\mu} \gamma_{\rho} L \nu_e )$  &
   $| \varepsilon^{eP}_{ {\mu}
  e}| <  5 \times 10^{-4}$&
\\  &$( \mu \rightarrow  3 e )\as$ & \\
  \hline
 $(\bar{u} \gamma^{\rho} P {u} )
   (\bar{\nu}_{\mu} \gamma_{\rho} L \nu_e )$  &
$| \varepsilon^{uP}_{ {\mu}
  e}| < 7.7\times 10^{-4} $& \\
&$({\rm Ti} \mu \rightarrow {\rm  Ti} e )\as$ & \\\hline
 $(\bar{d} \gamma^{\rho} P {d} )
   (\bar{\nu}_{\mu} \gamma_{\rho} L \nu_e )$ &
$| \varepsilon^{dP}_{ \mu e
 }| < 7.7\times 10^{-4} $& \\
  &$({\rm Ti} \mu \rightarrow  {\rm Ti} e )\as$ & \\
   \hline
%
%
 $(\bar{e} \gamma^{\rho} P {e} )
   (\bar{\nu}_{\tau} \gamma_{\rho} L \nu_e )$  &
  $| \varepsilon^{eP}_{ {\tau}
  e}| < 2.9 $  &
 $|  \varepsilon^{eL}_{{\tau}e
  }| < 0.02, |  \varepsilon^{eR}_{ {\tau} e }| < 0.04   $ \\
 &  $( \tau \rightarrow  e \bar{e} e )\as$ & leptonic $s_W^2$  at nufact \\
 &$|  \varepsilon^{eL}_{{\tau}e
  }| < 0.4, |  \varepsilon^{eR}_{ {\tau} e }| < 0.7   $ & \\
 &  LSND & \\ 
\hline
 $(\bar{u} \gamma^{\rho} P {u} )
   (\bar{\nu}_{\tau} \gamma_{\rho} L \nu_e )$  &
   $| \varepsilon^{uP}_{ {\tau}
  e}| < 1.6 $ &
 $|  \varepsilon^{uP}_{{\tau}
  e}| < 0.03   $
 \\ &$( \tau \rightarrow  e \rho)\as$
    & $s_W^2$ in DIS at nufact \\  
  &$|  \varepsilon^{uP}_{{\tau}e
  }| < 0.5  $ & \\
 &  CHARM& \\ \hline
 $(\bar{d} \gamma^{\rho} P {d} )
   (\bar{\nu}_{\tau} \gamma_{\rho} L \nu_e )$ &
   $| \varepsilon^{dP}_{\tau e}
 | < 1.6 $ &
 $|  \varepsilon^{dP}_{{\tau}
  e}| < 0.03   $
 \\ &$( \tau \rightarrow  e \rho)\as$
    & $s_W^2$ in DIS at nufact \\ 
  &$|  \varepsilon^{dP}_{{\tau}e
  }| < 0.5  $ & \\
 &  CHARM& \\
 \hline
  \hline
\end{tabular}

\end{table}
\renewcommand{\arraystretch}{1}

\begin{center}
{\Large {\bf Erratum to Present and future bounds on non-standard neutrino interactions}}
\end{center}

The renormalisation group mixing between operators that induce Non-Standard neutrino Interactions, such as 
$(\overline{\nu}_\tau \gamma^\rho P_L \nu_\mu)(\overline{e} \gamma_\rho e)$, and operators that induce flavour changing 
rare charged lepton decays(such as $\tau \to \mu \bar{e} e$), is not present at leading log. This is contrary to the claims of 
Section 3.2 and we thank Mattias Blennow, Carla Biggio and Enrique Fernandez-Martinez \cite{Blennow} 
for drawing our attention  to this error.
  
The bounds on non-standard neutrino Interactions (NSI) are affected as follows: there are finite contributions of the same order\footnote{in dimensional regularisation, and unitary gauge. There are additional diagrams in other gauges, but we obtain the same finite part.}, 
but unlike the leading log, the corrections are renormalisation scheme-dependent. The scheme dependence can be cancelled 
by matching corrections at the high scale, which are model dependent. Model and scheme independent bounds could be obtained by
performing a NLO calculation. 

\section{Details}

Recall that the NSI operators in which we are interested are
of  dimension 8, are $SU(3) \times SU(2) \times U(1)$ invariant,
and are constructed out of four fermion  and two  Higgs
fields.  A list of the relevant dimension 8 operators can be found in \cite{anna}.  
The operators are constructed such that
when the Higgs gets a vev, they induce the vertex
\beq
(\bar{f} \gamma^\rho f) (\bar{\nu}_\alpha \gamma_\rho \nu_\beta) ~~,
\label{vertex}
\eeq
but not
\beq
(\bar{f} \gamma^\rho f) (\bar{e}_\alpha \gamma_\rho e_\beta) ~~~~~,~~~~~
(\bar{f} \gamma^\rho f') (\bar{\nu}_\alpha \gamma_\rho e_\beta)~,
\label{not}
\eeq
where $f$ is one of $u,d$ or $e_R$, and $\alpha, \beta$ are
flavour indices. To simplify our calculation, we neglect
the possibility that $f = e_L$, which implies that  $W$  bosons
can not  be exchanged between the currents.

We want to obtain  the contribution 
to  charged lepton  vertices such as Eq.~\ref{not},
from the the  combination of electroweak
interactions and the New Physics that
induces   NSI. We opt to approach this problem from
a ``hard core'' Effective Field Theory approach.
We imagine the  calculation  in three steps \cite{BurasHouches}:
\begin{enumerate}
\item At the mass scale $\Lambda_{NSI}$ of the  New Physics
inducing NSI, match the full theory onto
the Standard Model (SM) plus effective  operators invariant
under SM gauge symmetries. This determines the (model-dependent)
dimensionless coefficients $\{ C_i \} $  
of the effective dimension 8  operators. 

 We assume that the coefficients
of  operators which
induce the undesirable vertices (\ref{not}) are  found to be zero
to all orders in the new interactions that generate them.
 This assumption gives the weakest bounds on NSI.   
It needs a significant amount of fine tuning to be arranged;
see the discussion in Ref.~\cite{Blennow}.
It could in principle arise due to a symmetry.

\item Run this effective theory  to $m_W$. 
\item Match at $m_W$ onto the $SU(3) \times U(1)$
theory (where there are no $W$s); there can be
calculable, finite terms from the matching at $m_W$. 
\end{enumerate}
The first step we do not consider further.  We are interested in
the effective theory calculation, which starts at step two, with
the boundary conditions we have assumed in step one.

We do not consider quadratic divergences.  This is because we
work in the Effective Field Theory formalism  where one calculates
in dimensional regularisation, runs with massless RGEs, and
at the  mass scale of each particle, matches the higher scale theory 
onto the lower scale theory (from which the particle has been removed).
In this formalism, quadratic divergences appear in the matching,
and by  assumption in step one above, they vanish.  

\subsection{ Running and matching in  broken SU(2)}

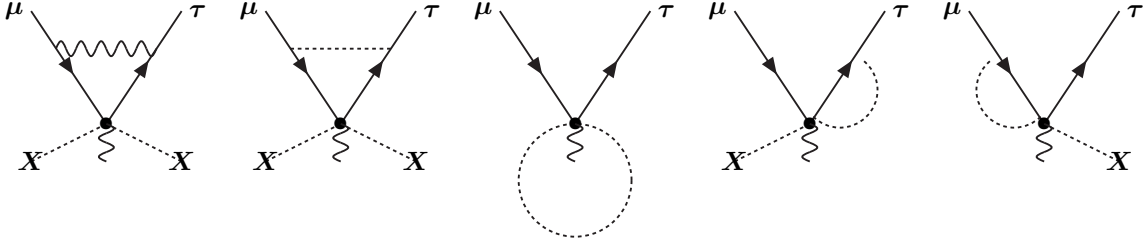
\begin{figure}[ht]
\unitlength.5mm
\SetScale{1.418}
\begin{boldmath}
\begin{center}
\vspace{1cm}
\begin{picture}(60,40)(0,0)
\ArrowLine(0,50)(20,20)
\ArrowLine(20,20)(40,50)
\Photon(6.6,40)(33.3,40){2}{5}
\DashLine(20,20)(40,10){1}
\DashLine(20,20)(0,10){1}
\Text(-2,50)[r]{$\mu$}
\Text(42,50)[l]{$\tau$}
\Text(40,10)[c]{$X$}
\Text(0,10)[c]{$X$}
\Text(20,20)[c]{$\bullet$}
\Photon(20,20)(20,10){2}{2}
\end{picture}
\begin{picture}(60,40)(0,0)
\ArrowLine(0,50)(20,20)
\ArrowLine(20,20)(40,50)
\DashLine(6.6,40)(33.3,40){1}
\DashLine(20,20)(40,10){1}
\DashLine(20,20)(0,10){1}
\Text(-2,50)[r]{$\mu$}
\Text(42,50)[l]{$\tau$}
\Text(40,10)[c]{$X$}
\Text(0,10)[c]{$X$}
\Text(20,20)[c]{$\bullet$}
\Photon(20,20)(20,10){2}{2}
\end{picture}
\begin{picture}(60,40)(0,0)
\ArrowLine(0,50)(20,20)
\ArrowLine(20,20)(40,50)
\DashCArc(20,5)(15,0,360){1}
\Text(-2,50)[r]{$\mu$}
\Text(42,50)[l]{$\tau$}
\Text(20,20)[c]{$\bullet$}
\Photon(20,20)(20,10){2}{2}
\end{picture}
\begin{picture}(60,40)(0,0)
\ArrowLine(0,50)(20,20)
\ArrowLine(20,20)(40,50)
\DashCArc(28,29)(10,225,50){1}
\DashLine(20,20)(0,10){1}
\Text(-2,50)[r]{$\mu$}
\Text(42,50)[l]{$\tau$}
\Text(0,10)[c]{$X$}
\Text(20,20)[c]{$\bullet$}
\Photon(20,20)(20,10){2}{2}
\end{picture}
\begin{picture}(60,40)(0,0)
\ArrowLine(0,50)(20,20)
\ArrowLine(20,20)(40,50)
\DashLine(20,20)(40,10){1}
\DashCArc(12,29)(10,130,315){1}
\Text(-2,50)[r]{$\mu$}
\Text(42,50)[l]{$\tau$}
\Text(40,10)[c]{$X$}
\Text(20,20)[c]{$\bullet$}
\Photon(20,20)(20,10){2}{2}
\end{picture}
\end{center}
\end{boldmath}
\caption{Relevant diagrams for the broken SU(2) calculation.
\label{figbroken}}
\end{figure}

To be concrete, suppose that  the fundamental theory
generates an  operator which  gives 
the vertex (\ref{vertex}), for $f = e_R$, with coefficient
$C_{NSI}$.  In unitary gauge, 
one can obtain  the first vertex of eqn (\ref{not}) by $W$ exchange
between the neutrino lines (the diagram on the left of
figure \ref{figbroken}). It is straightforward to verify,
using dimensional regularisation, that this diagram is finite. 
At one loop there is no log, in the lepton massless limit, so no mixing  during Renormalisation
Group running (step two above)  of the neutrino
vertex of  eqn (\ref{vertex}) into the charged lepton 
vertex of  eqn  (\ref{not}). 
In our original paper, we incorrectly guessed that this
anomalous dimension was non-zero so the log was present. 
However, the diagram has a finite part: at one-loop
matching at $m_W$,  the first vertex Eq.~\ref{not} is induced
with coefficient
\beq
- \frac{3m_W^2}{32 \pi^2} \frac{C_{NSI}}{\Lambda_{NSI}^4}~.
\label{finite}
\eeq
Unfortunately, this finite part is {\it not} scheme-independent. 
This problem is discussed in Section 1.3. 

We use Naive Dimensional Regularisation for $\gamma_5$.
We have checked that we get the same result calculating
the complete set of diagrams of Fig.~\ref{figbroken}  in
Landau gauge, and in a general $R-\xi$ gauge (where the gauge
dependence vanishes as it should).

\subsection{ Running and matching in  unbroken SU(2)}

In our EFT formalism, where particles are massless
when present in the theory, we could imagine SU(2) to be
unbroken between $\Lambda_{NSI} $ and $m_W$.  This perspective
allows a    simple, SU(2) invariant reformulation of
the previous calculation. 
Consider  the operator
\beq
 (\bar{f} \gamma^\rho  f)  J_\rho^{NSI}~, 
\label{simple}
\eeq
with 
\beq
 J_\rho^{NSI}  =  (\bar{\ell}_\alpha H ) \gamma_\rho (H^\dagger \ell_\beta) ~.
\label{JNSI}
\eeq
Recall $\alpha$, $\beta$ are flavour indices, $H$ is the Higgs doublet, $\ell_\alpha$ are the lepton doublets 
and  we use the diagonal SU(2) contraction
\beq
(H^\dagger \ell) = \left(  \begin{array}{cc}
H_{0}^* & 
H_{-}^* \end{array} \right)
\left(  \begin{array}{c}
\nu \\
e \end{array} \right)
 =  H^*_{0} \nu + H^*_{-} e~,
\eeq
where $\nu$ and $e$ are lefthanded fields.
 
Interactions among four charged leptons will be generated by Eq.~\ref{simple} but with $J^{NSI}_\rho$ replaced by
\beq
 J_\rho^{CLI}  =  (\bar{\ell}_\alpha \tilde{H} ) \gamma_\rho (\tilde{H}^\dagger \ell_\beta) ~,
\label{JCLI}
\eeq
with $\tilde{H}=i\tau_2 H^*$ and $i\tau_2$ is the antisymmetric SU(2) tensor.

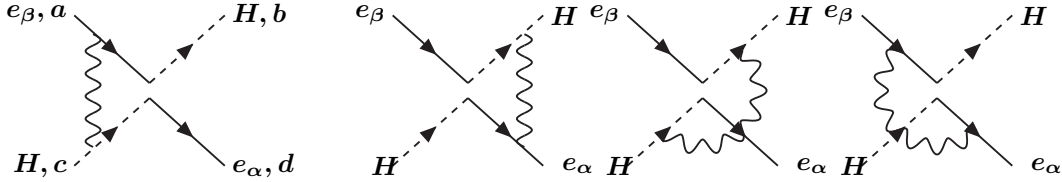
\begin{figure}[ht]
\unitlength.5mm
\SetScale{1.418}
\begin{boldmath}
\begin{center}
\begin{picture}(60,40)(0,0)
\ArrowLine(0,40)(20,22)
\DashArrowLine(20,22)(40,40){2}
\ArrowLine(20,18)(40,0)
\DashArrowLine(0,0)(20,18){2}
\Photon(5,35)(5,5){2}{5}
\Text(-2,40)[r]{$e_\b, a$}
\Text(42,40)[l]{$H,b$}
\Text(42,0)[l]{$e_\a,d$}
\Text(-2,0)[r]{$H, c$}
\end{picture}
\hspace{1cm}
\begin{picture}(60,40)(0,0)
\ArrowLine(0,40)(20,22)
\DashArrowLine(20,22)(40,40){2}
\ArrowLine(20,18)(40,0)
\DashArrowLine(0,0)(20,18){2}
\Photon(35,35)(35,5){2}{5}
\Text(-2,40)[r]{$e_\b$}
\Text(42,40)[l]{$H$}
\Text(45,0)[l]{$e_\a$}
\Text(-2,0)[c]{$H$}
\end{picture}
\begin{picture}(60,40)(0,0)
\ArrowLine(0,40)(20,22)
\DashArrowLine(20,22)(40,40){2}
\ArrowLine(20,18)(40,0)
\DashArrowLine(0,0)(20,18){2}
\PhotonArc(20,20)(15,225,45){2}{7}
\Text(-2,40)[r]{$e_\b$}
\Text(42,40)[l]{$H$}
\Text(45,0)[l]{$e_\a$}
\Text(-2,0)[c]{$H$}
\end{picture}
\begin{picture}(60,40)(0,0)
\ArrowLine(0,40)(20,22)
\DashArrowLine(20,22)(40,40){2}
\ArrowLine(20,18)(40,0)
\DashArrowLine(0,0)(20,18){2}
\PhotonArc(20,20)(15,135,315){2}{7}
\Text(-2,40)[r]{$e_\b$}
\Text(42,40)[l]{$H$}
\Text(45,0)[l]{$e_\a$}
\Text(-2,0)[c]{$H$}
\end{picture}
\end{center}
\end{boldmath}
\caption{\label{figunbroken} Current mixing diagrams for an unbroken SU(2).
$a,b,c,d$ are SU(2) indices, same for all diagrams.}
\end{figure}

We are then interested in the mixing of  $J_\mu^{NSI}$ 
into the charged lepton interaction current  $J_\mu^{CLI}$. 
We compute the one-loop electroweak corrections
to this mixing for an unbroken SU(2) theory containing
gauge bosons and the Higgs doublet. The relevant diagrams are in Fig.~\ref{figunbroken}.
We have used dimensional regularization and performed the calculation in a general 
gauge, checking the cancellation of the dependencies in the gauge parameter. 
The wave function renormalisation does not contribute to the mixing.
We find there are no logarithmic corrections of NSI into CLI, only a finite term 
which agrees with the result shown in Sec. 1.1 in the theory after SSB.

Alternatively, we can discuss the mixing between triplet and singlet currents

\beq
 J_\mu^{S} =  H^\dagger H 
\overline{\ell}_\alpha 
 \gamma_{\mu} \ell_{\beta}
~~~~~
J_\mu^{T} = \sum_A ~ H^\dagger \tau^A H 
~\overline{\ell}_\a 
\tau^A \gamma_\mu \ell_\b~,
\label{Jcl}
\eeq
where $\{ \tau^A \}$ are the SU(2) Pauli matrices.

NSI and CLI currents are simply related to singlet and triplet currents by

\beq
J_\mu^{NSI}=\frac{1}{2}J_\mu^S+\frac{1}{2}J_\mu^T~,~~~~
J_\mu^{CLI}=\frac{1}{2}J_\mu^S-\frac{1}{2}J_\mu^T .
\eeq

In this case, after inclusion of wave function renormalisation we find that:

\begin{itemize}
\item Singlet and triplet currents do not mix at all.
\item Singlet and and triplet currents, at one loop, have the same anomalous dimension.
\item In naive dimensional regularisation singlet and triplet currents obtain different finite contributions which once translated into the NSI and CLI currents reproduce the finite contribution shown before. 
\end{itemize}

\subsection{Renormalisation scheme dependence}
\label{section 1.3}

If one performs renormalisation group resummation in step two of our EFT prescription,  it
is well-known \cite{BurasHouches}  that  one-loop finite corrections   are of  the same order  
(in  the SU(2) coupling $\alpha_2$) as the two-loop
anomalous dimensions.  
Therefore,   to obtain a renormalisation-scheme
independent result, we could compute the  coefficient of the logarithm arising in the two-loop
mixing of $J^{NSI}$ into $J^S$. 
A useful discussion of the two-loop  QCD  anomalous dimension for a four-quark operator (similar to our four-leg operator, 
but with scalars replaced by fermions), using
various prescriptions for $\gamma_5$, can be found in
\cite{BurasWeisz}.

The SU(2) gauge coupling runs more slowly than $\alpha_{QCD}$,
so one may wonder whether it is neccessary to perform
RG resummation. Without it, 
the two-loop anomalous dimension is ${\cal O} (\alpha_2^2)$ 
and can be neglected. The complete calculation at given order in $\alpha_2$ should be renormalisation scheme independent, if we do perturbation theory  
without resummation (notice that we discuss renormalisation scheme dependence, not scale dependence).
{\it And}, the complete calculation  at ${\cal O}(\alpha_2)$
includes one loop matching, at the high scale $\Lambda_{NSI}$, of the
full theory onto the effective theory.  These
matching coefficients are clearly model-dependent. 
So if we do not use the EFT prescription of matching-resummation-matching,
then the  finite bit we have  obtained  must be summed with the  (model-dependent) 
finite bits from the high scale matching. Only if we resum,  do the scheme-dependent
 ${\cal O} (\alpha_2(m_W))$ terms cancel  between the  finite part and the anomalous dimension.
This   would allow a result which is formally independent of the high-scale theory.

\end{thebibliography}

\end{document}